\documentclass{jaa}
\usepackage{graphicx}
\usepackage{xcolor}
\newcommand{\msun}{\ensuremath{\rm M_{\odot}}}
\newcommand{\lum}{erg\,s$^{-1}$}
\newcommand{\fermi}{{\it Fermi}}
\newcommand{\nustar}{{\it NuSTAR}}
\newcommand{\xmm}{{\it XMM-Newton}}

\newcommand{\swift}{{\it Swift}}
\newcommand{\phflux}{\mbox{${\rm \, ph \,\, cm^{-2} \, s^{-1}}$}}

\newcommand{\gm}{$\gamma$}

\begin{document}\sloppy

%%paper title
%%For line breaks \\ can be used within title 
\title{Gamma-ray Emitting Narrow-Line Seyfert 1 Galaxies: Past, Present, and Future}

%%author names are separated by comma (,) 
%%use \and before the last author name 
%%use a * along with the number separated by comma
%% for the  author for correspondence
%%\textsuperscript{number} is used for affiliation
%%\affilOne, \affilTwo etc., upto \affilTwentyfive is possible
%%Please note the first letter after \affil is capitalised in the command
%%

\author{Vaidehi S. Paliya}
\affilOne{Deutsches Elektronen Synchrotron DESY, Platanenallee 6, D-15738 Zeuthen, Germany.\\}

%%escape two column mode for title, affiliation and abstract
%%by giving \twocolumn command as shown

\twocolumn[{

\maketitle

%%include \corres to print the corresponding author Email id
\corres{vaidehi.s.paliya@gmail.com}

%%include \msinfo for
%%manuscript information such as
%%received, revised and accepted dates
%%
\msinfo{7 July 2019}{2 September 2019}

%%abstract
\begin{abstract}
This article reviews our current understanding about \gm-ray detected narrow-line Seyfert 1 (\gm-NLSy1) galaxies. The detection with the Large Area Telescope onboard \fermi~Gamma-ray Space Telescope has provided the strongest evidence for the presence of closely aligned relativistic jet in these intriguing active galactic nuclei (AGN) and opened up a realm to explore the physical conditions needed to launch the jet in a different central engine and host galaxy environment than that is known for blazars. Promising results acquired from various multi-wavelength campaigns are converging to a scenario in which the \gm-NLSy1 galaxies can be considered as `young' blazars. These enigmatic sources hold the key to unravel the jet triggering mechanism and evolution of the AGN phase of a galaxy, in general. As such, \gm-NLSy1s should be considered as one of the top priority targets for next generation observational facilities.
\end{abstract}

%%insert keywords separated by 3 hyphens using \keywords{words}

}]
%%close the twocolumn escape here

%%include \doinum{number}for the DOI number in the header
%%include \volnum{number} for the volume number in the header
%%include \year{yyyy} for  year of publication in the header
%%include \pgrange{num--num} page range of article in the header
%%include \artcitid{num} for the article citation id
%%include \lp to print last page of the article
%%include \setcounter{page}{pagenum} for the exact starting page of the article

\doinum{12.3456/s78910-011-012-3}
\artcitid{\#\#\#\#}
\volnum{000}
\year{0000}
\pgrange{1--}
\setcounter{page}{1}
\lp{1}

\section{Introduction}\label{sec:intro}
\begin{figure*}[t]
\hbox{
\includegraphics[scale=0.56]{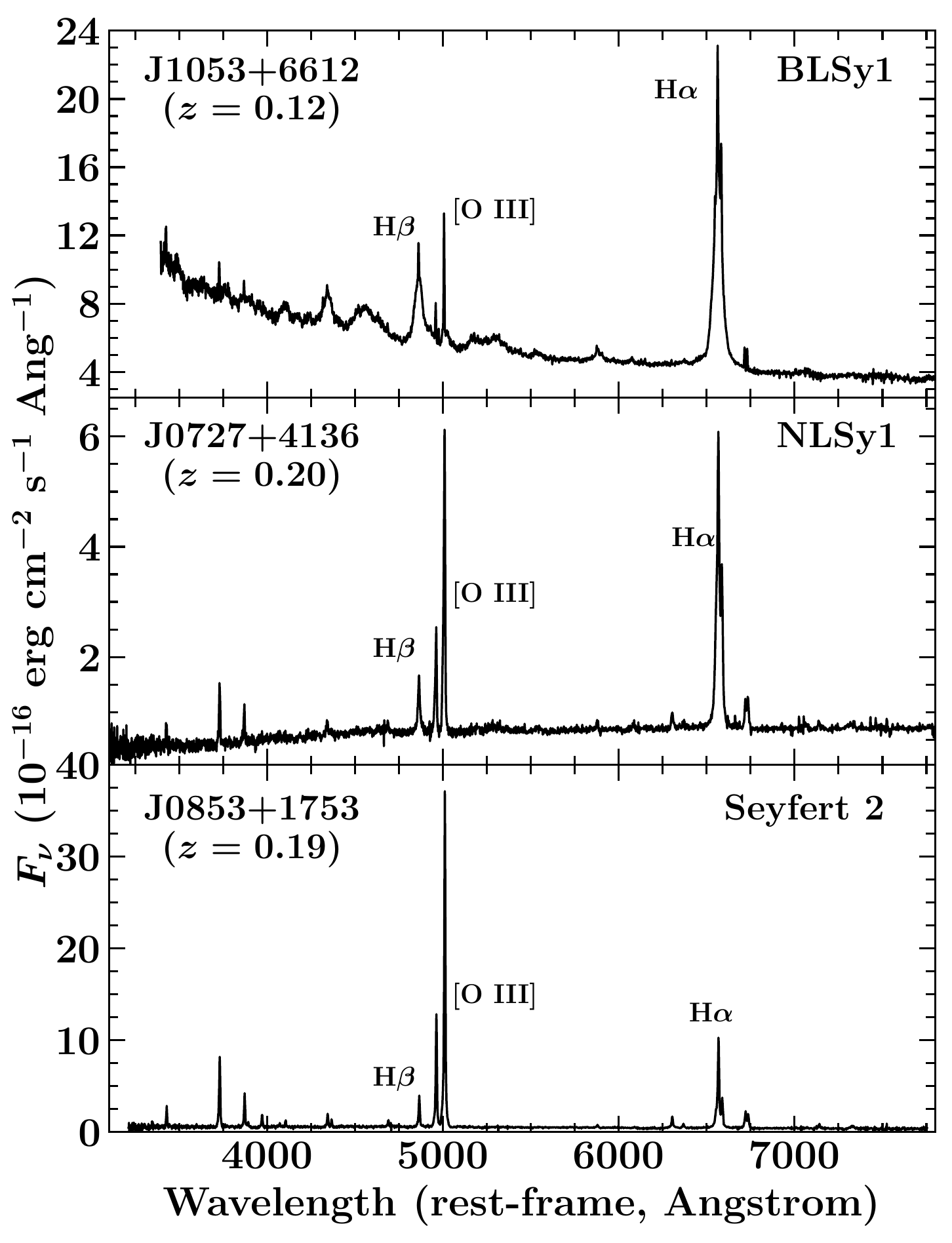}
\includegraphics[scale=0.56]{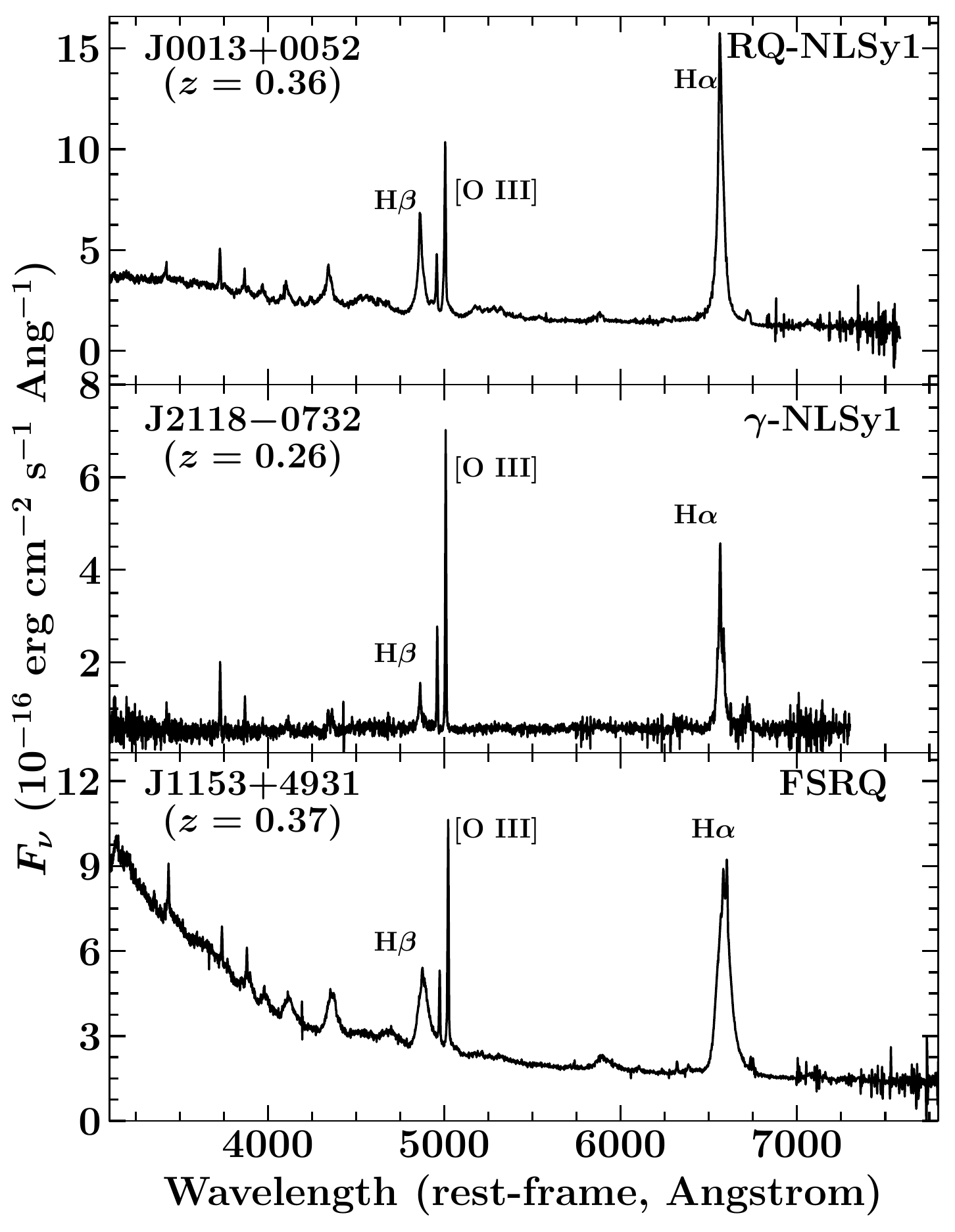}
}
\caption{Left: Rest-frame optical spectra of the broad-line Seyfert 1 (top), NLSy1 (middle), and Seyfert 2 (bottom) galaxies taken with SDSS. Right: Rest-frame optical spectra of a radio-quiet (top) and a \gm-ray loud (which is also radio-loud, middle) NLSy1, and a \fermi-LAT detected FSRQ (bottom). The most prominent emission lines are labeled.\label{fig:spec}}
\end{figure*}

Seyfert galaxies are low-luminosity active galactic nuclei (AGN, absolute magnitude $M_{\rm B}>-23$, e.g., V{\'e}ron-Cetty \& V{\'e}ron 2010). They are categorized as Seyfert 1 and Seyfert 2 (see, Netzer 2015, for a recent review) based on the presence or absence of broad permitted emission lines in their optical spectra, respectively (Figure~\ref{fig:spec}). The observational differences in these two classes can be explained by the AGN unification scheme (Antonucci 1993) with Type 2 sources viewed edge-on leading to the obscuration of the broad line emitting clouds by the molecular torus. Interestingly, some of the Seyfert 1 galaxies exhibit broad permitted lines with narrow width. Such AGNs are called narrow-line Seyfert 1 (NLSy1) galaxies (see Figure~\ref{fig:spec}). The defining criteria are full width at half-maximum or FWHM of ${\rm H\beta}<2000$ km s$^{-1}$, weak [O~III] emission line ([O~III]/H$\beta$ flux ratio $<$3), and the presence of strong Fe {\sc ii} multiplets (Osterbrock \& Pogge 1985; Goodrich 1989; Mullaney \& Ward 2008). The threshold of the FWHM of H$\beta$ emission line as 2000 km s$^{-1}$ appears a bit arbitrary as it does not reflect any sharp differences in the physical properties of AGNs around this value. A more conservative choice of 4000 km s$^{-1}$ have been used in many works (e.g., Sulentic {\em et al.} 2000; Marziani {\em et al.} 2018). This threshold differentiates the population A and B sources on the first fundamental correlation vector or the Eigenvector-1\footnote{The first eigenvector EV1 reflects the anti-correlations between [O~III]$\lambda$5007 peak intensity and FWHM of ${\rm H\beta}$ with FeII$\lambda$4570 (Boroson \& Green 1992).}) of type 1 AGNs (Boroson \& Green 1992).  In this plane, population A refers mainly to low black hole mass/high accretion rate sources and vice-versa for population B objects. Most of strongly radio-loud sources belong to population B. However, the original criterion of 2000 km s$^{-1}$ is adopted in most of the NLSy1 studies. In addition to that, by studying  296 NLSy1s from the Sloan Digital Sky Survey data release 7 (SDSS DR7), Cracco {\em et al.} (2016) have argued that a strong Fe {\sc ii} emission may not be a distinctive property of NLSy1s. 

Since the classification of an AGN as a NLSy1 object is based on the optical spectral properties, the sample of known NLSy1s has grown in the last two decades mainly due to advancement of the large optical spectroscopic surveys, mainly SDSS. The latest catalog of NLSy1 galaxies contains 11,101 sources based on SDSS  DR12 (Rakshit {\em et al.} 2017) which is about a factor of 5 improvement with respect to the previous catalog based on SDSS DR3 (Zhou {\em et al.} 2006). Moreover, using the 6-degree field galaxy survey data, Chen {\em et al.} (2018) reported the identification of 167 NLSy1 in the southern hemisphere. 

NLSy1s are X-ray bright sources, exhibit strong variability (e.g., Boller {\em et al.} 1996; Leighly 1999a; Komossa \& Meerschweinchen 2000), and generally show a steep X-ray spectrum (photon index $\Gamma_{\rm X}>2$, $\frac{dN}{dE}\propto E^{-\Gamma_{\rm X}}$, Grupe {\em et al.} 1998; Leighly 1999b). A prominent soft X-ray excess and a reflection dominated hard X-ray emission has also been observed from many NLSy1 galaxies (e.g., Fabian {\em et al.} 2009) revealing the characteristics of the X-ray corona and its interaction with the accretion disk. In fact, observations of NLSy1 galaxies from \xmm, \nustar, and other sensitive X-ray telescopes have enabled astronomers to explore the behavior of the matter and energy at the innermost region of the central engine (cf. Done \& Nayakshin 2007; Parker {\em et al.} 2014; Wilkins {\em et al.} 2015; Fabian {\em et al.} 2015, 2017; Kara {\em et al.} 2017). At ultraviolet (UV)/extreme UV frequencies, many NLSy1 objects exhibit a prominent big-blue-bump that is likely to originate from the accretion disk. Optical spectra of many of these sources exhibit strong outflows in the form of [O~III] blueshifts (Phillips 1976, Zamanov {\em et al.} 2002, Marziani {\em et al.} 2003, Bian, Yuan \& Zhao 2005, Komossa {\em et al.} 2008, Berton {\em et al.} 2016, Marziani {\em et al.} 2016, Komossa {\em et al.} 2018). These extreme physical properties of NLSy1s can be understood in terms of near-Eddington accretion onto low-mass black holes ($M_{\rm BH}\sim10^{6-8}$ \msun, Boroson 2002; Grupe \& Mathur 2004; Xu {\em et al.} 2012).

At GHz frequencies, more than 90\% NLSy1 galaxies are radio-quiet (radio-loudness index\footnote{The radio-loudness parameter $R$ is defined as the ratio of the flux density at 5 GHz over that at 4400 \AA~(Kellermann {\em et al.} 1989)). Since radio-loudness of an AGN is directly connected with the presence/absence of the jet, $R$ is often used to identify jetted AGNs. A few recent studies, however, have suggested that connecting presence/absence of the jet with the radio-loudness can be misleading (e.g., Padovani 2017, L{\"a}hteenm{\"a}ki {\em et al.} 2018).} $R<10$, Ulvestad {\em et al.} 1995; Moran 2000; Boroson 2002; Komossa {\em et al.} 2006; Singh \& Chand 2018). This is in contrast to the quasar population where $\gtrsim$15\% sources are found to be radio-loud (Kellermann {\em et al.} 1989, 2016). 

A significant boost to the studies of NLSy1 galaxies came with the \gm-ray detection from 4 radio-loud NLSy1s within the first year of \fermi~Large Area Telescope (LAT) operation (Abdo {\em et al.} 2009b,c). Since \gm-ray detection indicates some of the most exotic physical processes at work, these \gm-ray emitting NLSy1 (\gm-NLSy1 hereafter) sources caught the attention of the astronomy community. Particularly, \fermi-LAT observations confirmed the presence of closely aligned relativistic jets in \gm-NLSy1s similar to blazars which are radio-loud quasars hosting powerful jets pointed towards the observer (e.g., Urry \& Padovani 1995). Due to peculiar orientation of blazar jets, the radiation originated from a region moving with bulk Lorentz factor $\Gamma \equiv (1 - \beta^2)^{-1/2}$, where $\beta$c is the jet speed, along the jet, is Doppler boosted in frequency by a factor $\delta = \left( \Gamma [ 1 - \beta \cos\theta_{\rm v} ] \right)^{-1}$, and in luminosity by a factor $\delta^4$ with respect to quantities measured in the intrinsic emission region frame. This phenomena makes blazars one of the very few astrophysical sources which are observable across the whole electromagnetic spectrum, i.e., from radio-to-very high-energy \gm-rays. Other defining features of blazars are the large amplitude, rapid flux and spectral variability (cf. Gaidos {\em et al.} 1996; Stalin {\em et al.} 2004; Aleksi{\'c} {\em et al.} 2011; Gopal-Krishna {\em et al.} 2011; Fuhrmann {\em et al.} 2014; Paliya {\em et al.} 2015a; Marscher 2016), flat or inverted radio spectrum ($\alpha>-0.5, F_{\nu}\propto\nu^{\alpha}$), high-brightness temperature ($>10^{10-11}$ K), and superluminal motion (see, e.g., Scheuer \& Readhead 1979; Jorstad {\em et al.} 2005; Pushkarev {\em et al.} 2009; Homan {\em et al.} 2015; Jorstad {\em et al.} 2017; Piner \& Edwards 2018; Lister {\em et al.} 2019). Based on the rest-frame equivalent width (EW) of the broad emission lines, blazars are classified as flat spectrum radio quasars (FSRQs, EW$>$5\AA) and BL Lac objects (EW$<$5\AA, Stickel {\em et al.} 1991).

This article reviews our current knowledge about the physical characteristics of the known \gm-NLSy1 galaxies, their similarity/dissimilarity with blazars, and how can these intriguing beamed systems be used to probe the jet launching/triggering processes and the evolution of relativistic jets. The basic overview of the blazar-like properties exhibited by NLSy1 class from a historical, pre-\fermi~($<$2008) perspective is presented in Section~\ref{sec:past} The lessons learned from various extensive multi-frequency campaigns organized in the \fermi~era (2008$-$), are briefly reported in Section~\ref{sec:pres} Main science problems associated with jet physics that can be addressed with the observations of \gm-NLSy1 galaxies from the current/next generation observational facilities are discussed in Section~\ref{sec:fut} Finally, summary of the reported results is presented in Section~\ref{sec:con}

\section{Past: NLSy1 as Candidate Beamed AGNs}\label{sec:past}
Till the beginning of the twenty-first century, most of the NLSy1 research was focused on their optical and X-ray characteristics and only a handful of them were studied in the radio domain (e.g., Remillard {\em et al.} 1986; Grupe {\em et al.} 2000; Oshlack {\em et al.} 2001; Zhou {\em et al.} 2003) mainly due to their low radio-loud fraction. A systematic search for radio-loud NLSy1s was done by Komossa {\em et al.} (2006) leading to the identification of a total of 11 objects. A majority of them exhibit steep radio spectrum and do not show indications of beaming (see also Whalen {\em et al.} 2006). On the other hand, studies focused on individual NLSy1 sources reported a few peculiar ones, e.g., PMN J0948+0022 ($z=0.59$) shows radio flux variations, very high radio brightness temperatures ($>10^{13}$ K, Zhou {\em et al.} 2003), indicating relativistically beamed emission (see also Wang {\em et al.} 2006). Zhou {\em et al.} (2005) argued SBS 0846+513 ($z=0.58$) as a hybrid NLSy1-blazar source. 

The most robust arguments to support the presence of closely aligned relativistic jets in a radio-loud NLSy1 galaxy, 1H 0323+342 ($z=0.06$), were presented by Zhou {\em et al.} (2007). The optical spectrum of this object fulfills the usual definition of a NLSy1, whereas, it also exhibits multi-frequency behavior that is characteristics of blazars. This includes a flat radio spectrum at GHz frequencies and a compact core plus one-sided jet structure on milliarcsecond scale at 8.4 GHz resolved with Very Long Baseline Array (VLBA). The source is highly variable at the radio, optical, and X-ray energies. It was also observed from Whipple telescope to search for high-energy \gm-ray emission above 400 GeV, though the detection significance was marginal (Falcone {\em et al.} 2004).

The growing evidence about some radio-loud NLSy1s exhibiting blazar-type features were systematically explored by Yuan {\em et al.} (2008) (see also Foschini {\em et al.} 2009). From a sample of 2011 NLSy1s present in SDSS-DR3 (Zhou {\em et al.} 2006), they selected 23 extremely radio-loud objects with $R>100$. Most of them exhibit flat radio spectra, high brightness temperature and blue optical slopes indicating a significant contribution from the non-thermal `jet' component. The derived radio-to-optical and optical-to-X-ray slopes ($\alpha_{\rm ro}$ and $\alpha_{\rm ox}$, respectively) were consistent with that known from blazars. In the X-ray band, many radio-loud NLSy1 galaxies exhibit a rising spectrum in the $\nu F_{\nu}$ versus $\nu$ plane ($\Gamma_{\rm X}\lesssim1.8$) which can be interpreted as due to inverse Compton process. These observational characteristics led Yuan {\em et al.} (2008) to conclude that all radio-loud NLSy1 objects host small-scale, mildly relativistic jets similar to  classical radio-loud quasars. Flat radio spectrum NLSy1s are postulated as the beamed sources, whereas, compact steep spectrum NLSy1 objects probably belong to the parent population, i.e., sources with jets viewed at large viewing angles with respect to the line of sight. These speculations were soon going to be tested with the \fermi-LAT which was scheduled to be launched in 2008 June.

\section{Present: Low-Luminosity Blazars}\label{sec:pres}
\begin{table*}
\tabularfont
\begin{center}
\caption{List of \gm-ray detected NLSy1 galaxies.\label{tab:basic_info}}
\begin{tabular}{lcccccl}
\hline
Name & R. A. (J2000) & Decl. (J2000) & $F_{\rm 1.4~GHz}$ & $g'$ & $z$  &  References \\
 & hh mm ss.s & dd mm ss & (mJy) & (mag) &  & \\
\hline
	                               &                    &  Flat spectrum                   &               &         &         &                 \\
	                               &                    &  \gm-NLSy1 galaxies                   &               &         &         &                 \\
1H 0323+342	              & 03 24 41.1 & +34 10 46   & 613.5     & 15.7 & 0.60 & Abdo {\em et al.} (2009c)   \\
SBS 0846+513              & 08 49 58.0 & +51 08 29   & 266.3   & 18.8 & 0.58 & Foschini (2011)     \\
CGRaBS J0932+5306  & 09 32 41.1 & +53 06 33   & 481.6   & 18.9 & 0.60 & Paliya {\em et al.} (2018)   \\
GB6 J0937+5008$^*$ & 09 37 12.3 & +50 08 52   & 166.6   & 19.1 & 0.28 &  Paliya {\em et al.} (2018)  \\
TXS 0943+105$^*$      & 09 46 35.1 & +10 17 06.6 & 410.3  & 19.5 & 1.00  & Yao {\em et al.} (2019) \\
PMN J0948+0022        & 09 48 57.3 & +00 22 26    & 69.5   & 18.6 & 0.58 &  Abdo {\em et al.} (2009b)  \\
TXS 0955+326	          & 09 58 20.9 & +32 24 02    & 1247.1 & 16.1 & 0.53 &  Paliya {\em et al.} (2018)   \\
CGRaBS J1222+0413$^*$ & 12 22 22.5 & +04 13 16    & 800.3   & 17.0 & 0.97 & Yao {\em et al.} (20195b)  \\
SDSS J1246+0238       & 12 46 34.7 & +02 38 09 & 35.7 & 18.4 & 0.36         & Foschini (2011) \\
GB6 B1303+5132     & 13 05 22.9 & +51 16 39    & 87.0      & 17.3 & 0.79  &     Liao {\em et al.} (2015) \\
TXS 1419+391   	          & 14 21 06.0 & +38 55 23    & 85.7     & 18.6 & 0.49 &   Paliya {\em et al.} (2018) \\
GB6 B1441+4738     & 14 43 18.6 & +47 25 56   & 165.8     & 18.1  & 0.70 &    Liao {\em et al.} (2015)   \\
PKS 1502+036	          & 15 05 06.5 & +03 26 31    & 394.8   & 18.6 & 0.41 &   Abdo {\em et al.} (2009c)   \\
TXS 1518+423	$^*$      & 15 20 39.6 & +42 11 09    & 138.4   & 19.3 & 0.48      &  Paliya {\em et al.} (2018)  \\
RGB J1644+263          & 16 44 42.5 & +26 19 13    & 128.4   & 18.0 & 0.14  &    D'Ammando {\em et al.} (2015a)  \\
PKS 2004$-$447         & 20 07 55.2 & $-$44 34 44 & 471.0 & 19.3 & 0.24 &   Abdo {\em et al.} (2009c)  \\
TXS 2116$-$077          & 21 18 52.9  & $-$07 32 28 & 96.1       & 16.5  & 0.26 & Paliya {\em et al.} (2018)   \\
PMN J2118+0013$^*$ & 21 18 17.4 & +00 13 17    & 147.9   & 19.3 & 0.46       & Paliya {\em et al.} (2018)  \\
\hline
	                               &                    &  Steep spectrum                   &               &         &         &                 \\
	                               &                    &  \gm-NLSy1 galaxies                   &               &         &         &                 \\
FBQS J1102+2239        & 11 02 22.9  & +22 39 19 & 2.5      & 19.2 & 0.45 &      Foschini (2011)  \\
3C 286$^*$              & 13 31 08.3 & +30 30 32   & 14902.7 & 17.3 & 0.85 &     Berton {\em et al.} (2017) \\
\hline
	                               &                    &  Other                                           &               &         &         &                 \\
	                               &                    &  \gm-NLSy1 galaxies                   &               &         &         &                 \\
SDSS J0031+0936$^*$ & 00 31 59.9 & +09 36 18 & -- & 19.2 & 0.22 & Paiano {\em et al.} (2019) \\
SDSS J1641+3454          & 16 41 00.1  & +34 54 53 & 370 & 17.9 & 0.16  & L{\"a}hteenm{\"a}ki {\em et al.} (2018)\\
\hline
\end{tabular}

\end{center}
\tablenotes{Radio coordinates and 1.4 GHz flux values are adopted from NRAO VLA Sky Survey (Condon {\em et al.} 1998) and $g'$ magnitudes are from SDSS, except for PKS 2004$-$447, SDSS J003159.85+093618.4, and SDSS J164100.10+345452.7. The information about PKS 2004$-$447 are collected from Gallo {\em et al.} (2006). SDSS J003159.85+093618.4 is proposed as the counterpart of the unidentified \gm-ray source 3FGL J0031.6+0938 and its optical spectrum resemble well with NLSy1 galaxies (Paiano {\em et al.} 2019). This source has not been detected in NVSS or FIRST surveys, however, it is spatially coincident with a UV (GALEXASC J003159.86+093618.7) and an X-ray (XRT J003159+093615) object. The NLSy1 galaxy SDSS J164100.10+345452.7 ($z=0.16$) was reported as a \gm-ray emitter by L{\"a}hteenm{\"a}ki {\em et al.} (2018), however, it was not confirmed (Ciprini 2018). The \gm-NLSy1 galaxies GB6 B1303+5132 and GB6 B1441+4738 were reported as compact steep spectrum objects (Liao {\em et al.} 2015), however, Berton {\em et al.} (2018) reclassified them as flat spectrum NLSy1s. The same authors also proposed FBQS J1102+2239 as a compact steep spectrum source. Names with an asterisk are sources which were already known as \gm-ray emitters prior to their NLSy1 classification. TXS 1419+391 is a candidate NLSy1 source. Its H$\beta$ line profile in the SDSS optical spectrum is incomplete leading to the ambiguity in the line FWHM measurement (Rakshit {\em et al.} 2017). The last column are the \gm-ray discovery papers or where the association of the \gm-ray source with the NLSy1 was made.
}
\end{table*}

\begin{figure*}[t]
\hbox{
\includegraphics[width=\linewidth, trim=0 20 0 160, clip]{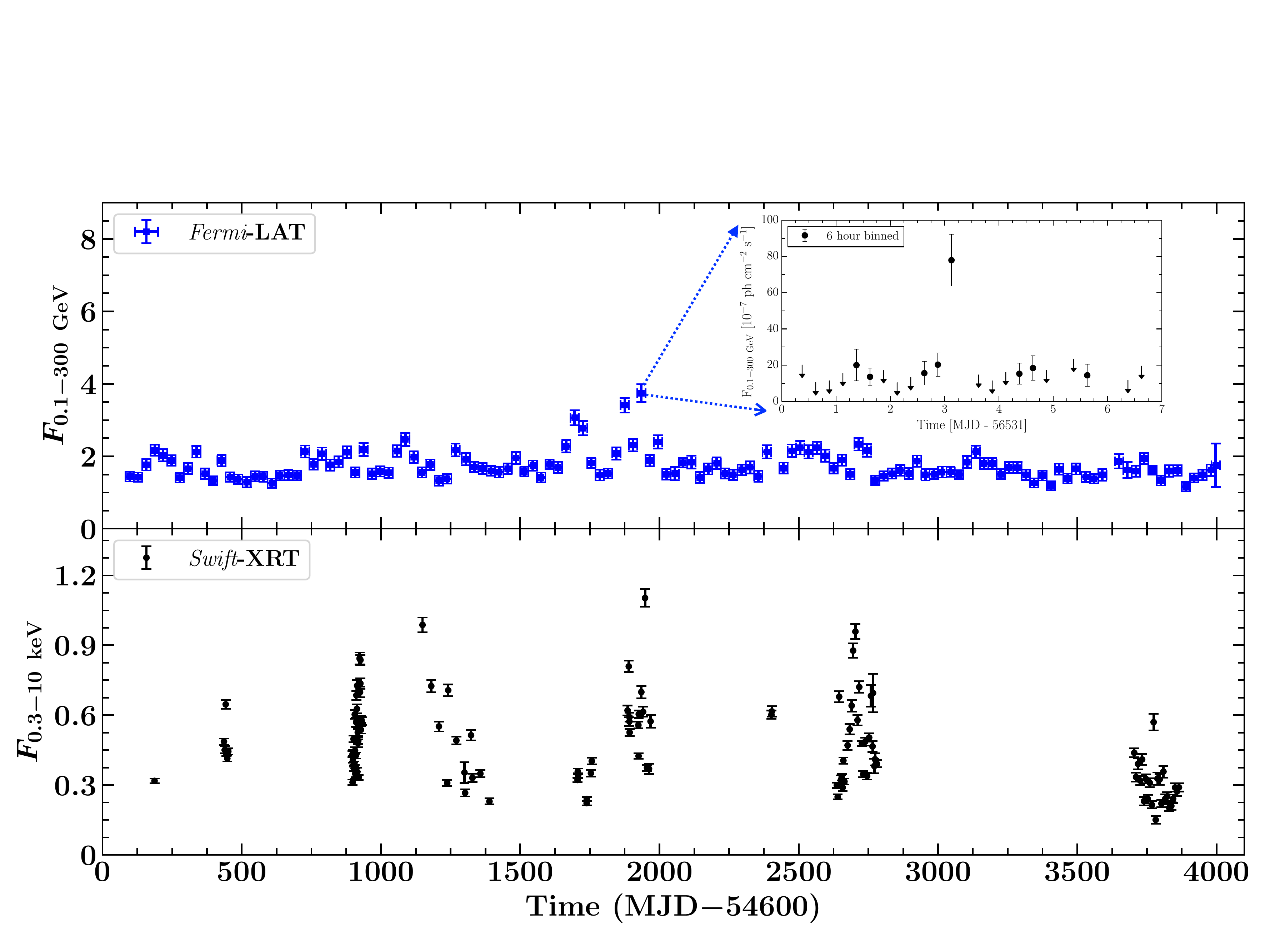}
}
\caption{Temporal behavior of the \gm-NLSy1 galaxy 1H 0323+342 in the \gm- and X-ray bands. Monthly binned \fermi-LAT data are taken from the public aperture photometry light curve repository (https://fermi.gsfc.nasa.gov/ssc/data/access/lat/4yr\_catalog/ap\_lcs.php) and has the unit of 10$^{-7}$ \phflux. \swift-X-ray Telescope (XRT) light curve is generated following Stroh \& Falcone (2013) in units of counts s$^{-1}$. Note the contrasting behavior of the source at \gm- and X-ray energies. The object exhibited a rapid \gm-ray variability during its 2013 September outburst (around MJD 56535, see inset). The downward arrows in inset refer to flux upper limits computed at 95\% confidence level. Inset is adopted from Paliya {\em et al.} (2015b) and used here with  permission from the AAS.\label{fig:1H323_g_lc}}
\end{figure*}

The \gm-ray detection of a flat spectrum, radio-loud AGN is probably the most convincing signature for the relativistically beamed emission. Blazars were known as the most numerous \gm-ray emitters in the extragalactic sky since Energetic Gamma-Ray Experiment Telescope (EGRET, Thompson {\em et al.} 1995) days and the same trend continued in the \fermi-LAT era. However, with a factor of $\sim$20 improvement in the sensitivity with respect to EGRET, \fermi-LAT was expected to detect \gm-ray emission from a lot more variety of astrophysical sources.

Within the first six months of \fermi-LAT operation, a \gm-ray source positionally consistent with the radio-loud NLSy1 galaxy PMN J0948+0022 was found (Abdo {\em et al.} 2009b). By integrating the first year of the \fermi-LAT data, three more NLSy1s (1H 0323+342, PKS 1502+036, and PKS 2004$-$447) were detected in the \gm-ray band with $>$5$\sigma$ significance (Abdo {\em et al.} 2009c). Later, by utilizing larger \fermi-LAT data set, a few more radio-loud NLSy1 galaxies were detected at \gm-ray energies (Foschini 2011; D'Ammando {\em et al.} 2015a; Paliya {\em et al.} 2018). In addition to the dedicated \gm-NLSy1 searches using the \fermi-LAT data, a few known \gm-ray emitting blazars were found to be NLSy1 sources when their optical spectra were carefully analyzed (Yao {\em et al.} 2015b; Paliya {\em et al.} 2018; Yao {\em et al.} 2019). Table~\ref{tab:basic_info} lists all of the known \gm-NLSy1 galaxies as of now.

Attempts were also made to search for \gm-ray emission from compact steep spectrum (CSS) NLSy1 galaxies. These objects are proposed as the members of the parent population of the beamed NLSy1s (Berton {\em et al.} 2015). Indeed, significant \gm-ray emission has been detected from 2 CSS NLSy1 galaxies (see Table~\ref{tab:basic_info}). 

Though the presence of relativistic jets in radio-loud NLSy1 galaxies was speculated even before the launch of the \fermi~satellite, the \gm-ray detection was a major breakthrough which fully established the fact that they do host closely aligned relativistic jets. However, this discovery raised questions about the physical condition needed to launch jets. Blazars, in general, are found to be hosted in giant elliptical galaxies (cf. Olgu{\'{\i}}n-Iglesias {\em et al.} 2016) and powered by massive black holes ($M_{\rm BH}>10^8$ \msun, Shaw {\em et al.} 2012). On the other hand, NLSy1 systems are usually hosted in spirals with low mass black holes ($M_{\rm BH}\sim10^{6-8}$ \msun) at their centers (cf. Krongold {\em et al.} 2001, Crenshaw {\em et al.} 2003, Grupe \& Mathur 2004; Deo {\em et al.} 2006; Ohta {\em et al.} 2007). A number of multi-frequency followup observations were carried out to address these outstanding issues and also to explore the similarity/dissimilarity of \gm-NLSy1 galaxies with blazars in a broadband context. A few important findings are listed below.

\subsection{Variability}
Flux variability is a defining feature of beamed AGNs. Blazars are known to display large amplitude variability, on all possible timescales, i.e., ranging from minutes to years, across the electromagnetic spectrum. Therefore, it is of a great interest to determine the flux variability behavior of \gm-NLSy1 galaxies. In fact, variability, if observed, can be used as an argument to exclude the origin of the \gm-ray emission due to star-forming activities (see, e.g., Ackermann {\em et al.} 2012) in NLSy1 sources.

\begin{figure*}[t]
\hbox{
\includegraphics[scale=0.4]{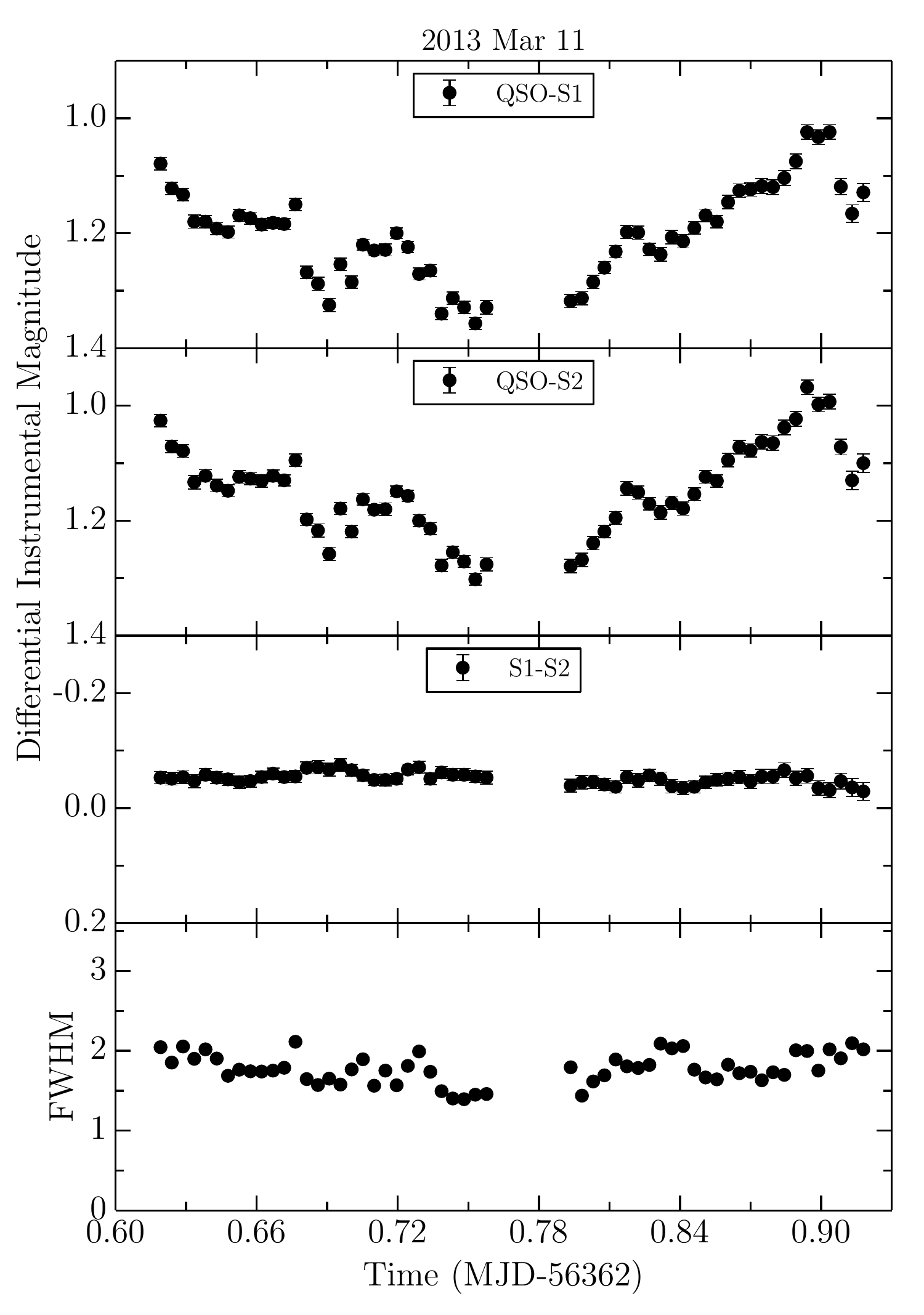}
\includegraphics[scale=0.5]{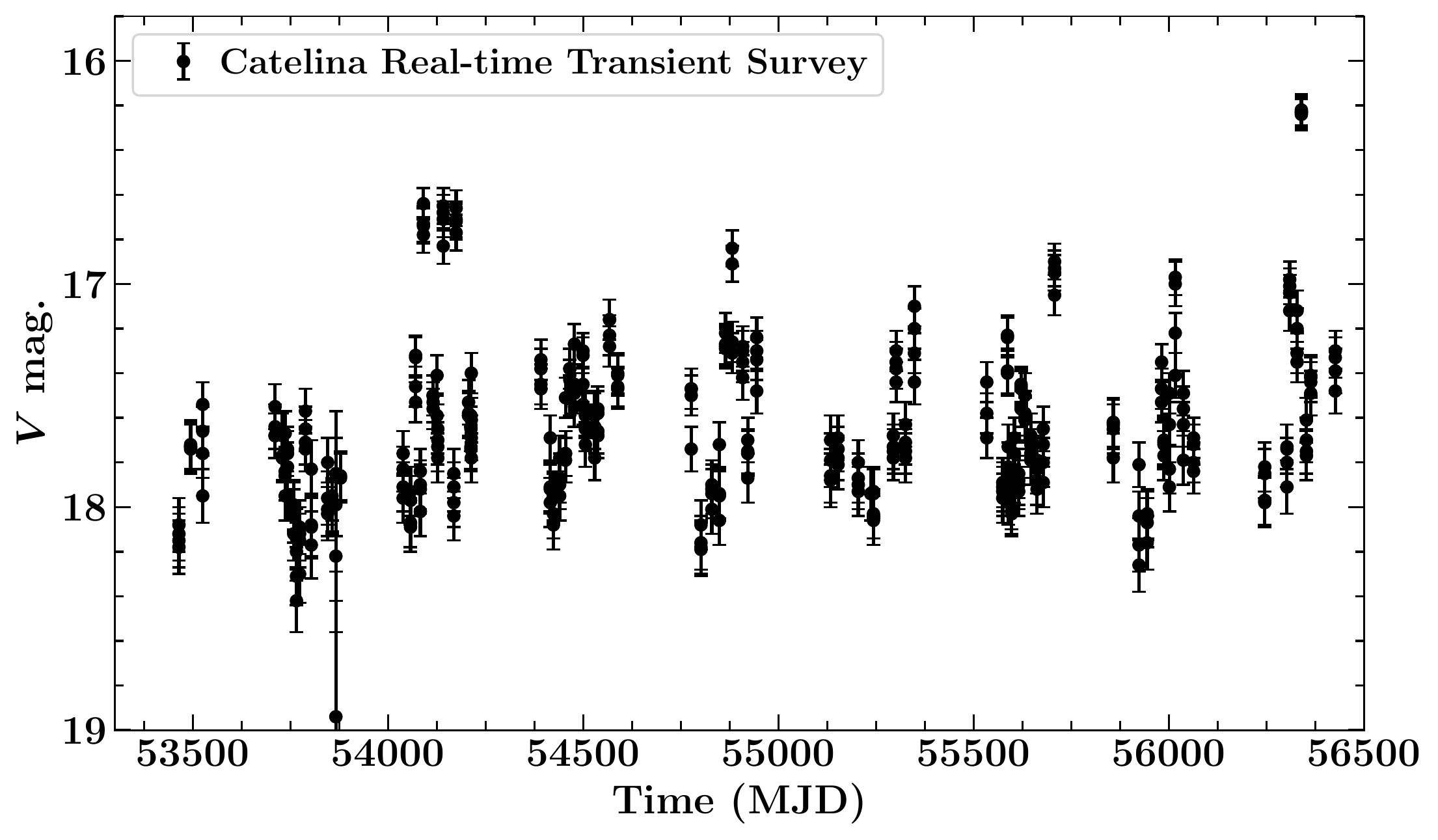}
}
\caption{Left: The $R$ band intra-night differential light curve of a \gm-NLSy1 galaxy, SBS 0846+513. The term `QSO' refers to the \gm-NLSy1 galaxy, whereas, `S1' and `S2' represent two comparison stars. The bottom panel shows the variation of the FWHM of the stellar images during the observing period. The plot is adopted from Paliya {\em et al.} (2016). Right: The long-term $V$ band light curve of a \gm-NLSy1, PMN J0948+0022. The data is adopted from the Catelina Real-time Transient Survey whose details can be found at Drake {\em et al.} (2009).}\label{fig:J0846_o_lc}
\end{figure*}

{\bf Gamma-rays:} It became evident soon after their \fermi-LAT detection that \gm-NLSy1 galaxies exhibit large amplitude GeV flares. The prototype of this class, PMN J0948+0022, exhibited the first \gm-ray outburst ever detected from a NLSy1 galaxy in 2010 when its daily binned \gm-ray flux, $F_{\gamma}$, exceeded $10^{-6}$ \phflux, corresponding to an isotropic \gm-ray luminosity $>10^{48}$ \lum~(Foschini {\em et al.} 2011, 2012). Later, a few more \gm-NLSy1s showed high activities in the \fermi-LAT energy range (D'Ammando {\em et al.} 2012; Paliya {\em et al.} 2014, 2015b; Paliya \& Stalin 2016). Interestingly, the 2013 September outburst of the nearest known \gm-NLSy1 source, 1H 0323+342, revealed a rapid \gm-ray flare with the shortest flux doubling time down to $\sim$3 hours (see Figure~\ref{fig:1H323_g_lc}, Paliya {\em et al.} 2014) indicating the \gm-ray emission to be originated from an extremely compact region in the jet. This kind of enigmatic \gm-ray behavior is typically observed from FSRQs class of blazars (e.g., Tavecchio {\em et al.} 2010; Tanaka {\em et al.} 2011; Abdo {\em et al.} 2011; Paliya 2015; Shukla {\em et al.} 2018). 
%On the other hand, since the \gm-NLSy1s are relatively faint sources, their duty cycle (i.e., fraction of time during which a source remain in an elevated activate state) is low.

{\bf X-rays:} Unlike the surveying operation of the \fermi-LAT, X-ray telescopes operate mostly in pointed mode, thus denying a continuous sky coverage to explore the X-ray variability behavior of \gm-NLSy1 objects. However, existing observations provide evidences about their intriguing nature. For example: a comparison of the \gm- and X-ray light curves of 1H 0323+342 led to the observation of a number of `orphan' X-ray flares with no \gm-ray counterparts (Figure~\ref{fig:1H323_g_lc}, see also Paliya {\em et al.} 2014). A possible explanation could be due to emission region being located very close to the central black hole where the absorption of the \gm-ray photons by \gm\gm~pair production with the bright X-ray corona emission can be important (e.g., Dondi \& Ghisellini 1995, Ghisellini \& Madau 1996). An observational signature of this process could be the detection of a soft spectrum in the X-ray band (see Ghisellini \& Tavecchio 2009, for a quantitative treatment) which is observed (Paliya {\em et al.} 2014). Other, rather sporadic, observations taken during the \gm-ray flaring episodes revealed a high level of 0.3$-$10 keV emission suggesting co-spatial origin of the observed radiations (e.g., Foschini {\em et al.} 2012; D'Ammando {\em et al.} 2015b; Paliya \& Stalin 2016). A few \gm-NLSy1s have also been targeted with the hard X-ray satellite \nustar, however, no significant flux variability has been found during the epoch of \nustar~observation (e.g., Landt {\em et al.} 2017). Since the hard X-ray emission in \gm-NLSy1 galaxies is found to be dominated by the jet radiation, the lack of variability could be due to low Doppler factor and/or the relatively large size of the emission region, assuming the causality argument ($R\lesssim \frac{c\Delta t\delta}{(1+z)}$, $\Delta t$ is the variability timescale) to be valid. Alternatively, the low jet activity state during the epoch of the \nustar~observations can also lead to the non-detection of rapid flux variations.

\begin{figure*}[t]
\hbox{
\includegraphics[width=\linewidth]{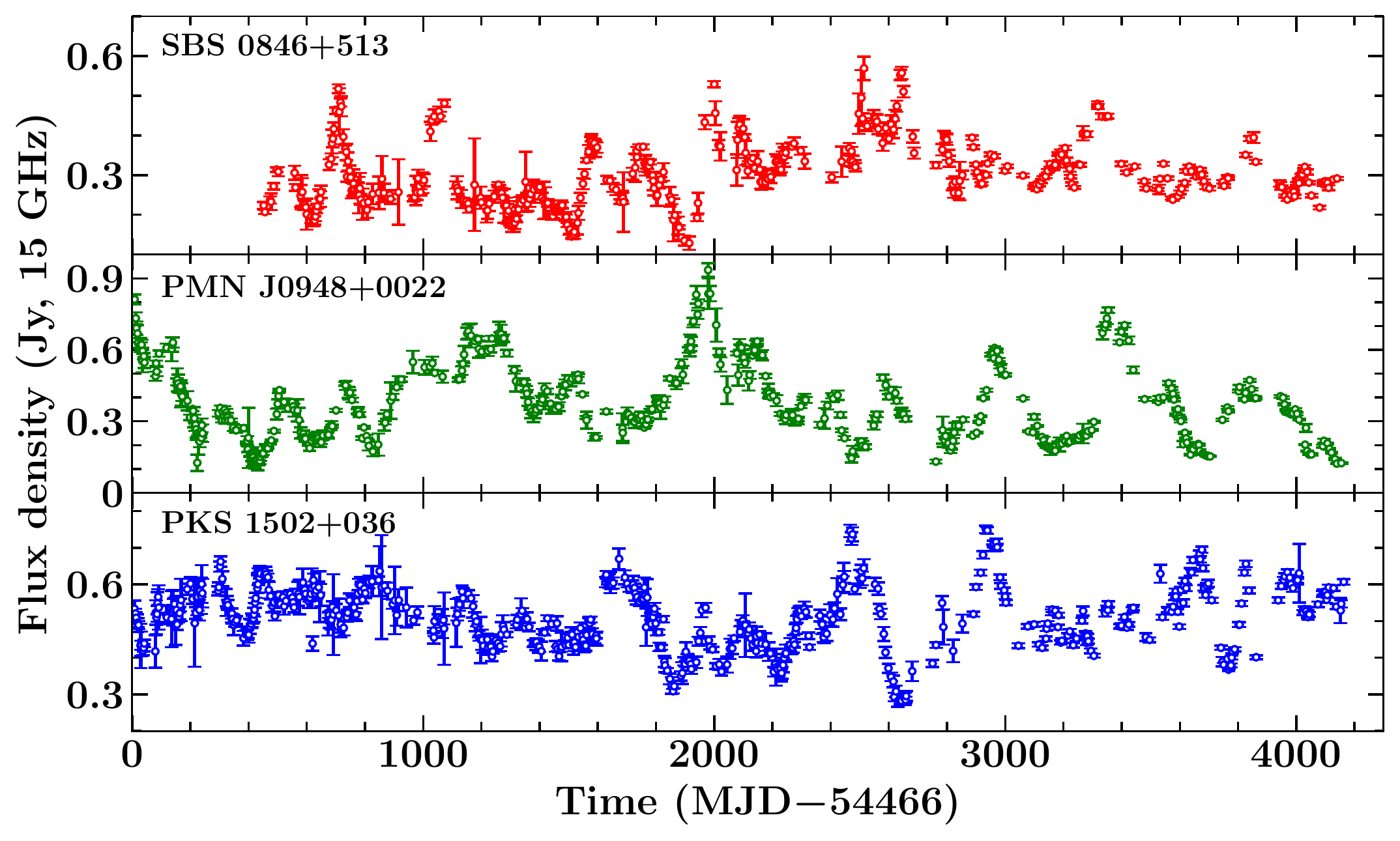}
}
\caption{15 GHz flux density measurements of three \gm-NLSy1 galaxies, as labeled, using the publicly available data from OVRO. The details of the data reduction procedure can be found at Richards {\em et al.} (2011).}\label{fig:ovro}
\end{figure*}

{\bf Optical:} At optical frequencies, \gm-NLSy1 galaxies exhibit violent flux variability as inferred both from the observations conducted during a single night and also at longer timescales ($\sim$months to years). Rapid optical flux variations  of $\gtrsim$0.3 magnitude within a few hours (i.e., Intra-Night Optical Variability or INOV) are reported for 1H 0323+342 (Paliya {\em et al.} 2014; Ojha {\em et al.} 2019), SBS 0846+513 (Maune {\em et al.} 2014; Paliya {\em et al.} 2016), and PMN J0948+0022 (Liu {\em et al.} 2010; Maune {\em et al.} 2013), including minute-scale optical flare detected from PMN J0948+0022 (Paliya {\em et al.} 2013a). These objects have also exhibited $>$1 magnitude flux variations on the timescale of a few months to years (Figure~\ref{fig:J0846_o_lc}, Maune {\em et al.} 2013, 2014). Interestingly, a significant INOV is found to coincide with the elevated \gm-ray activity state for a few \gm-NLSy1s (see Figure~\ref{fig:J0846_o_lc}, Paliya {\em et al.} 2014, 2016). These observations provide unambiguous evidences about the relativistically beamed synchrotron emission in \gm-NLSy1 sources similar to that observed from blazar class of AGNs (see, e.g., Sagar {\em et al.} 2004; Stalin {\em et al.} 2004; Gopal-Krishna {\em et al.} 2011; Paliya {\em et al.} 2017b). Furthermore, the duty cycle of the observed INOV in \gm-NLSy1s is found to be comparable or even higher to that noted for blazars (Paliya {\em et al.} 2013a; Kshama {\em et al.} 2017). A comparative study of the optical variability properties of different classes of NLSy1 galaxies has revealed \gm-NLSy1 to show higher duty cycle with respect to radio-quiet and radio-loud NLSy1s pointing towards the stronger beaming effect in \gm-ray detected objects (Kshama {\em et al.} 2017).

A few \gm-NLSy1 sources have also exhibited intriguing optical polarization behavior (e.g., Maune {\em et al.} 2014; Angelakis {\em et al.} 2018). In particular, polarized flux variability down to minute-scale was detected from PMN J0948+0022 during its 2012 December \gm-ray flare (Itoh {\em et al.} 2013). The detection of a high degree of polarization ($\sim$36$^{\circ}$) during this outburst indicates a highly ordered magnetic field inside the compact emission region. This kind of extreme behavior is more common in blazars (cf. Hagen-Thorn {\em et al.} 2008; Sasada {\em et al.} 2008) and is usually explained with multi-zone emission models (Marscher 2014).

{\bf Infrared:} The IR emission in radio-loud NLSy1 galaxies could originate from the star formation, dusty torus, or from the jetted synchrotron emission or all. NLSy1s, in general, are known to show enhanced star-formation activities as revealed by {\it Spitzer} and {\it Wide-field Infrared Survey Explorer} or {\it WISE} missions (Sani {\em et al.} 2010; Caccianiga {\em et al.} 2015). Moreover, radio detected NLSy1 objects are found to be more variable compared to their radio-quiet counterparts (Rakshit \& Stalin 2017). Considering \gm-NLSy1 galaxies, a significant IR variability is reported for a few sources (Jiang {\em et al.} 2012; Yao {\em et al.} 2015b; Yang {\em et al.} 2018; Gab{\'a}nyi, Moor \& Frey 2018, Yao {\em et al.} 2019). In particular, the rapid IR flux variations ($\bigtriangleup m\sim$0.1$-$0.2 mag within a day) constrained the size of the emission region as $<$10$^{-3}$ pc, following causality argument $R'\lesssim\frac{c\bigtriangleup t\delta}{(1+z)}$ where $\delta$ is the Doppler factor, thus effectively ruling out its origin from the significantly larger dusty torus (Jiang {\em et al.} 2012) or from the star-forming activities. These findings further strengthens the idea of the jet based origin of the IR flux variability in \gm-NLSy1s similar to that noted for blazars (Mao {\em et al.} 2018).

{\bf Radio:} Only a handful of \gm-NLSy1 galaxies have been regularly monitored from radio telescopes. Over a decade of 15 GHz flux density measurements acquired from Ovens Valley Radio Observatory (OVRO, Richards {\em et al.} 2011) have revealed multiple large amplitude flares (see Figure~\ref{fig:ovro}). The variability brightness temperature and associated Doppler factor derived from the multi-epoch, multi-frequency radio observations of some of the \gm-NLSy1s suggest the presence of mildly relativistic flows ($\delta\sim5-10$, Angelakis {\em et al.} 2015; Orienti {\em et al.} 2015; L{\"a}hteenm{\"a}ki {\em et al.} 2017). These findings support a close resemblance of \gm-NLSy1 objects with blazars. Interestingly, radio emission can also originate from the star-forming activities which has led to the observation of a tight correlation between IR and radio luminosities expected from starbursts (cf. Yun {\em et al.} 2001). However, many jetted NLSy1s are found to be significantly more radio luminous than expected from star-formation, hence indicating the jet based origin (Komossa {\em et al.} 2006). The radio emission in a few other NLSy1 objects, on the other hand, are found to be dominated by the star-formation related activities (Caccianiga {\em et al.} 2015).

\subsection{Spectral Energy Distribution}
An exhaustive study of the broadband spectral energy distribution (SED) enables us to explore the physical processes responsible for the observed emission from an astrophysical object. The radio-to-\gm-ray SED of a blazar consists of two broad humps. The low energy peak lying between sub-mm to X-rays bands is understood due to synchrotron process, whereas, the high-energy emission arises from the inverse Compton mechanism in the conventional leptonic emission scenario. Abdo {\em et al.} (2010b) introduced a classification scheme based on the location of the synchrotron peak frequency. Blazars whose rest-frame synchrotron peak lies below $\nu^{\rm syn}_{\rm peak,~Hz}=10^{14}$ are called low synchrotron peak or LSP sources. On the other hand, objects with $10^{14}\leqslant\nu^{\rm syn}_{\rm peak,~Hz}\leqslant10^{15}$ and $\nu^{\rm syn}_{\rm peak,~Hz}>10^{15}$ are termed as intermediate synchrotron peaked (ISP) and high-synchrotron peaked (HSP) blazars, respectively. The observed anti-correlation of the synchrotron peak luminosity and the peak frequency (Fossati {\em et al.} 1998) suggests that powerful blazars, i.e., FSRQs, are usually LSP sources, whereas, low-luminosity BL Lac objects mostly belong to ISP and HSP category. The shift of the synchrotron peak to lower frequencies enables the detection of the underlying accretion disk radiation in the form of the big-blue-bump in many FSRQs. In fact, since FSRQs exhibit broad optical emission lines indicating an efficient accretion process at work, the disk radiation cannot be much lower compared to the jet emission. 

In the conventional one-zone leptonic models, the electron population which emit synchrotron radiation, also participates in the inverse Compton mechanism suggesting both SED peaks to be connected. Accordingly, the high-energy peak in FSRQs lies at relatively lower, $\sim$MeV, energies, thereby making their X-ray spectrum to be flat rising and \gm-ray SED to be steep falling (in $\nu F\nu$ versus $\nu$ plane) which is confirmed from multi-wavelength observations (e.g., Abdo {\em et al.} 2010b). In contrast, inverse Compton peak in BL Lac sources is usually located at higher GeV$-$TeV energies, making their \gm-ray spectrum harder and often extends to TeV energy range\footnote{http://tevcat.uchicago.edu}. Moreover, the seed photons for the inverse Compton scattering could be different for various blazar types. The luminous accretion disk in powerful FSRQs illuminates the broad line region (BLR) and the dusty torus, thus a dense photon field surrounding the jet can interact with the relativistic jet electrons via so-called external Compton mechanism (see, e.g., Sikora {\em et al.} 1994; B{\l}a{\.z}ejowski {\em et al.} 2000). This process is usually invoked to explain the X- to \gm-ray SED observed in FSRQs (cf. Sikora {\em et al.} 2009). On the other hand, the high-energy emission in BL Lac objects  is typically explained via synchrotron self Compton or SSC mechanism where the synchrotron photons gets upscattered by the same electron population that emits them (e.g., Marscher \& Gear 1985).

The above mentioned description about blazar SED features in different energy bands and physical mechanisms powering the blazar jets can be used to understand their applicability and comparison with the \gm-NLSy1 galaxies. A brief discussion about the spectral behavior of these sources at different energies and also about their broadband SED is provided below. 

\begin{figure}[t]
\hbox{
\includegraphics[width=\linewidth]{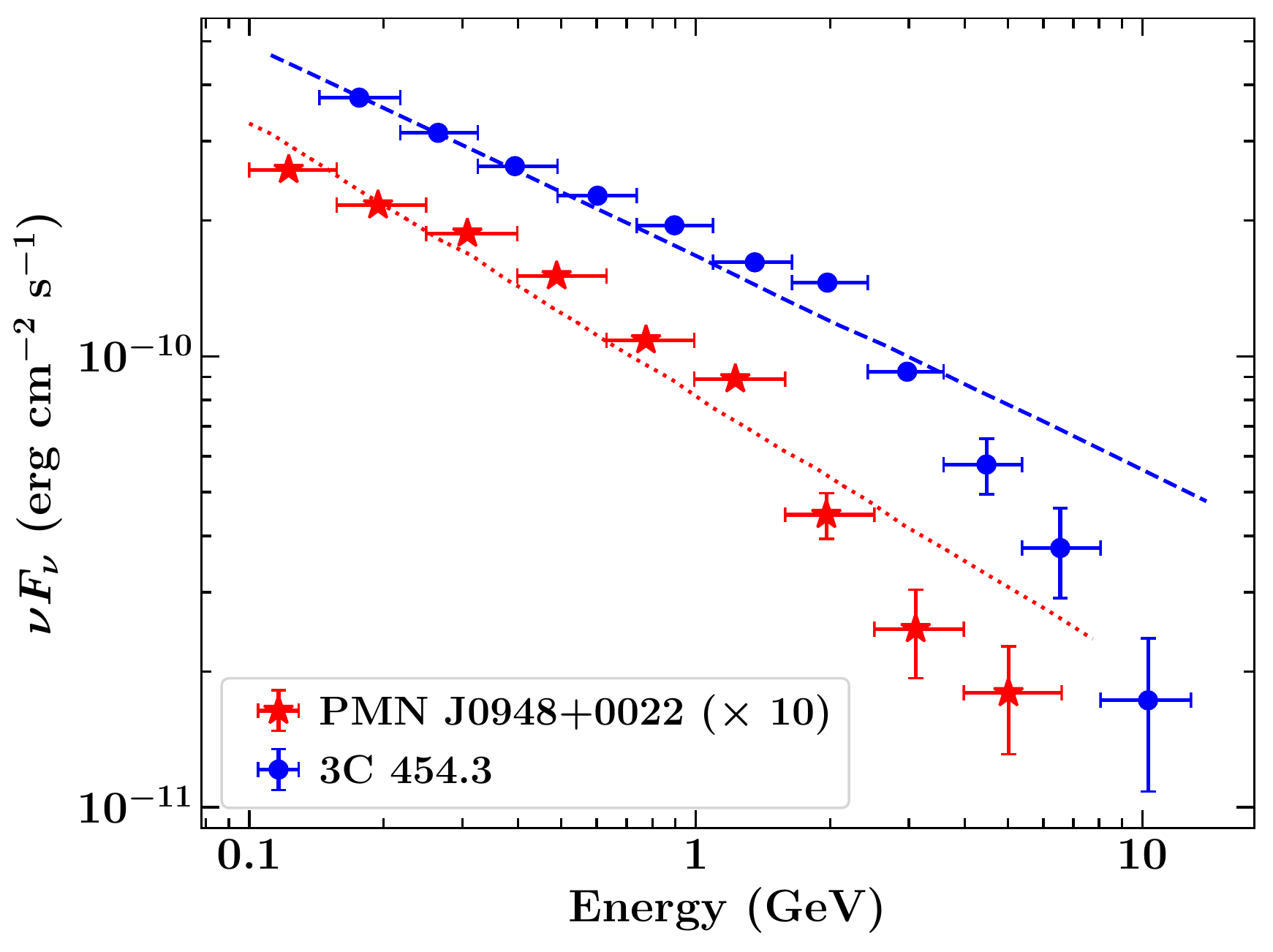}
}
\caption{Gamma-ray spectra of a \gm-NLSy1 galaxy, PMN J0948+0022 (red stars, multiplied by 10), and the well-known FSRQ 3C 454.3 (blue circles). The dotted and dashed lines represent the best-fitted power law models, respectively. Note the presence of a break or curvature around 1 GeV, indicating similar mechanisms at work in the relativistic jets of the \gm-NLSy1 and FSRQs. The data points are adopted from Paliya {\em et al.} (2015b) and Abdo {\em et al.} (2010a) for PMN J0948+0022 and 3C 454.3, respectively.}\label{fig:curve}
\end{figure}

{\bf Gamma-rays:} In the \gm-ray band accessible from the \fermi-LAT, all NLSy1 sources exhibit a steep falling spectrum (\gm-ray photon index $>$2, e.g., The Fermi-LAT collaboration 2019). This observation suggests their similarity with FSRQ class of blazars rather than with BL Lac objects. Interestingly, the \gm-ray spectra of a few NLSy1 galaxies show a pronounced curvature (Paliya {\em et al.} 2015b), a feature generally observed in powerful FSRQs (Abdo {\em et al.} 2009a, 2010a). An example is demonstrated in Figure~\ref{fig:curve}. Various theoretical models have been proposed to explain this feature. This includes absorption of high-energy \gm-ray photons within the BLR (Poutanen \& Stern 2010), Klein-Nishina effects (Cerruti {\em et al.} 2013), external Compton scattering of the hybrid photon field (Finke \& Dermer 2010), or intrinsic curvature in the emitting electron population (Abdo {\em et al.} 2009a). Though these hypotheses are still being debated, existing observations provide clear evidences that similar physical processes also operate in the jets of \gm-NLSy1s.

{\bf X-rays:} The X-ray spectrum of NLSy1 sources, and many radio-quiet AGNs in general, exhibit various intriguing observational features associated with the accretion disk-corona interaction, such as an excess emission below 2 keV, so-called soft X-ray excess (Arnaud {\em et al.} 1985, Singh, Garmire, and Nousek 1985), Fe K$\alpha$ line at 6.4 keV (Tanaka {\em et al.} 1995), and a Compton reflection hump at higher energies (cf. Fabian {\em et al.} 2004, 2009). These observational features possibly originate from the innermost region of the accretion disk and allows one to understand the mutual interplay of the disk and corona radiations and thus revealing the behavior of matter and energy at the vicinity of the central engine (e.g., Boller {\em et al.} 2003; Dewangan {\em et al.} 2007; Done \& Nayakshin 2007; Gallo {\em et al.} 2013; Fabian {\em et al.} 2015, 2017). According to the AGN unification scheme (e.g., Urry \& Padovani 1995), the X-ray emitting corona is expected to be present in all types of AGNs, however, only a minor fraction of them host relativistic jets, thereby making it tedious to study the disk-corona-jet connection. Additionally, in jetted AGNs like blazars, thermal coronal radiation is negligible compared to the Doppler boosted jet emission. This is the reason that, other than a few low-significance detections (e.g., Page {\em et al.} 2004; Madsen {\em et al.} 2015), a definitive coronal emission has never been observed from blazars. On the other hand, the \gm-NLSy1 galaxies host relatively lower power jets (see, e.g., Foschini {\em et al.} 2015) and hence it may be possible to detect and disentangle the X-ray corona emission.

\begin{figure}[t]
\hbox{
\includegraphics[width=\linewidth]{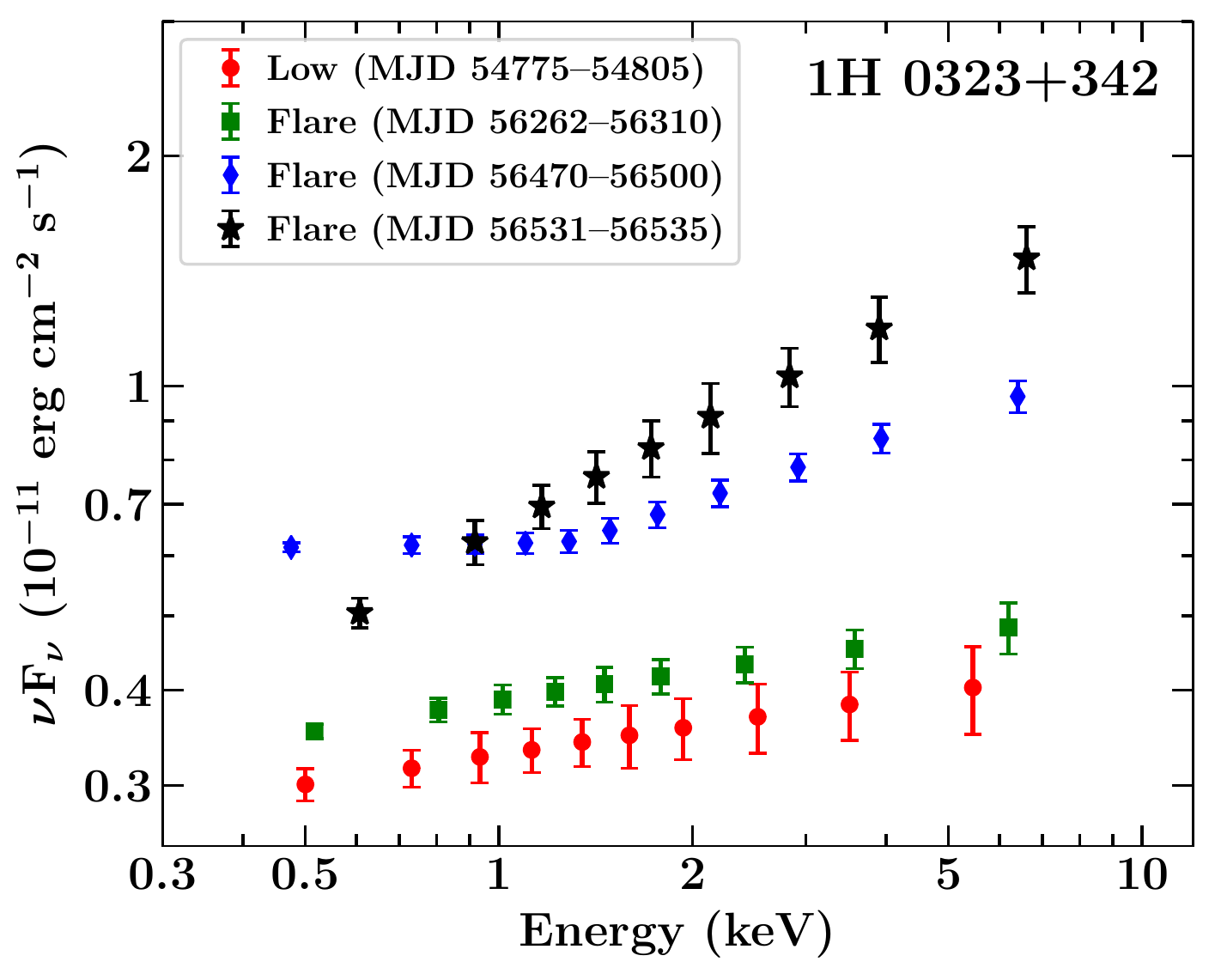}
}
\caption{The X-ray spectrum of 1H 0323+342 during various \gm-ray, or equivalently jet activity, states as labeled and studied in Paliya {\em et al.} (2014). The black stars refer to the 2013 September \gm-ray outburst of the source. Note the spectral hardening as the flux increases indicating the dominance of the jet emission.}\label{fig:evolution}
\end{figure}

The X-ray spectra of radio-loud NLSy1 galaxies is harder with respect to that observed from their radio-quiet counterpart ($\Gamma_{\rm X}\sim2$ and 2.5, respectively, Brandt {\em et al.} 1997; Grupe {\em et al.} 2010; Foschini {\em et al.} 2015). The \gm-NLSy1 galaxies, on average, also display similar X-ray spectral properties (Paliya {\em et al.} 2019a). Interestingly, the activity dependent spectral evolution is observed in one of the \gm-NLSy1 galaxy 1H 0323+342. The X-ray spectrum of this source is dominated by the soft coronal emission during quiescence (cf. Sun {\em et al.} 2015; Yao {\em et al.} 2015a; Ghosh {\em et al.} 2018). However, during elevated \gm-ray activity states, jet radiation dominates making the X-ray spectrum harder (see Figure~\ref{fig:evolution}, Paliya {\em et al.} 2014).

\begin{figure*}[t]
\hbox{
\includegraphics[scale=0.78]{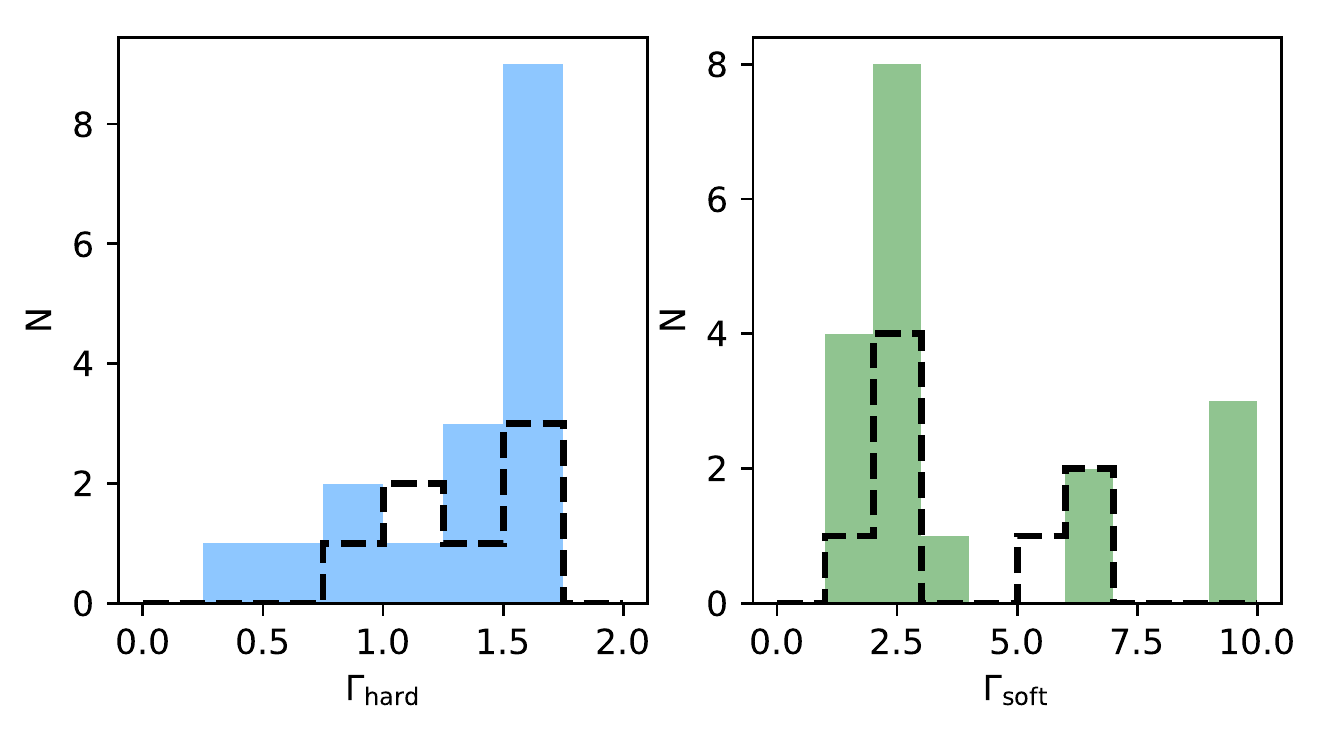}
\includegraphics[scale=0.4]{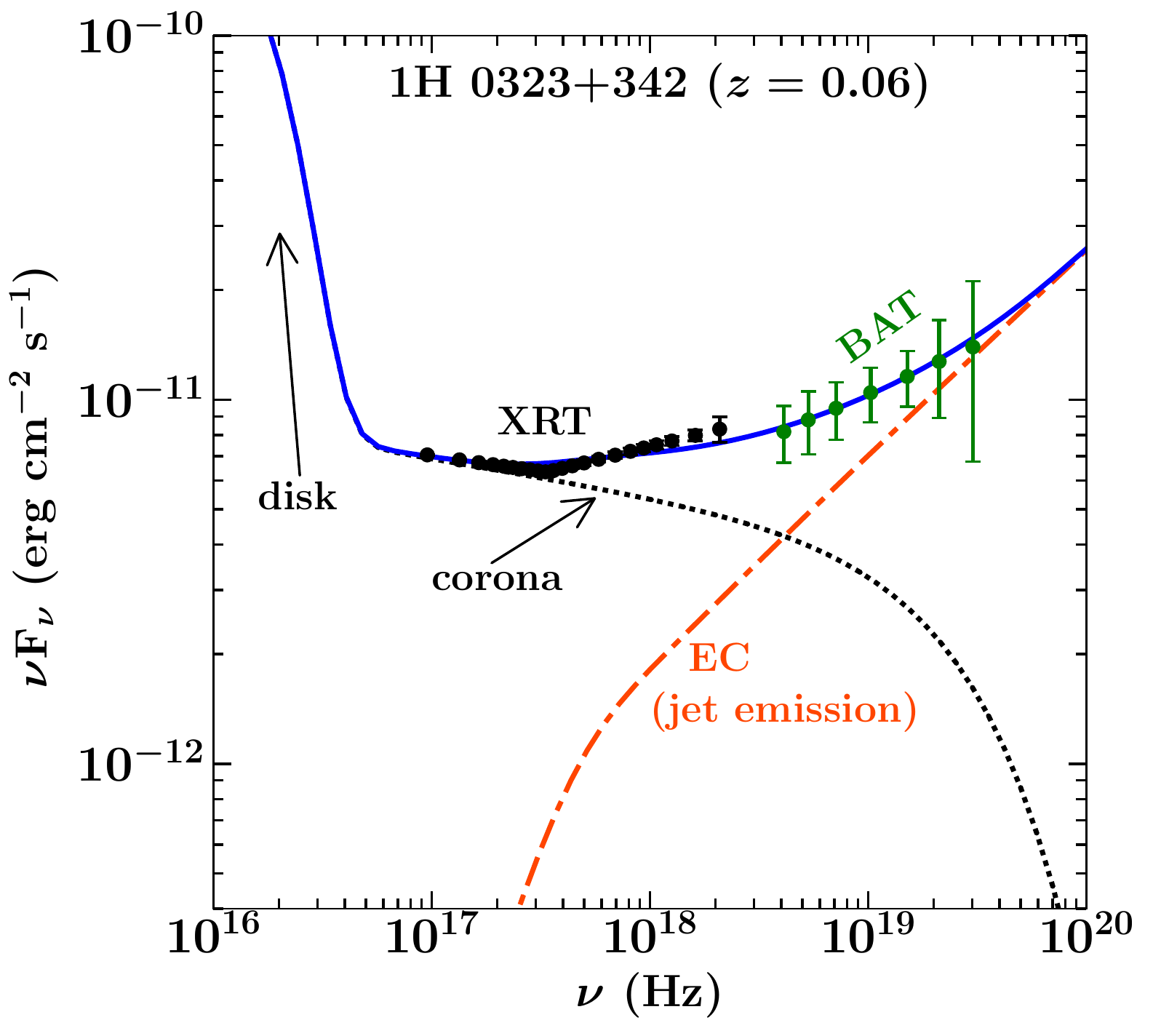}
}
\caption{Left: The distribution of two power law indices after ($\Gamma_{\rm hard}$) and before ($\Gamma_{\rm soft}$) the break energy in the X-ray spectra of \gm-NLSy1s. The plot is adopted from Paliya {\em et al.} (2019a). Right: An X-ray zoomed plot of the multi-wavelength SED of the \gm-NLSy1 galaxy 1H 0323+342 which is fitted with a leptonic jet model (see Paliya {\em et al.} 2019a). Note the break at a few keV above which the spectrum hardens suggesting the dominance of the external Compton or EC radiation from the jet. At lower energies, the spectral shape is pretty soft indicative of the corona emission.}\label{fig:break}
\end{figure*}

The X-ray spectra of many \gm-NLSy1s exhibit a break around $\sim$2 keV (see, e.g., Larsson {\em et al.} 2018; Paliya {\em et al.} 2019a) as derived by fitting a broken power law model. On average, the spectrum before the break is found to be soft ($\Gamma_{\rm X}\sim2.5$) probably due to coronal emission (soft X-ray excess), whereas, the jet radiation dominates at higher energies leading to the observation of a hard SED ($\Gamma_{\rm X}\sim1.6$). This is shown in Figure~\ref{fig:break} which also demonstrate that the \gm-NLSy1 sources are probably the only known jetted population whose X-ray spectra reveal both jet and corona radiation signatures.

A few \gm-NLSy1s also exhibit soft X-ray excess similar to radio-quiet NLSy1 galaxies (e.g., Bhattacharyya {\em et al.} 2014; Kynoch {\em et al.} 2018). Various models have been put forward to explain this feature such as thermal Comptonization of the accretion disk emission by the coronal plasma (Titarchuk 1994) and the relativistically blurred reflection model (cf. Ross \& Fabian 2005) However, since deep X-ray observations, e.g., with \xmm, are not available for a majority of \gm-NLSy1s, a strong remark about the nature of the soft X-ray excess cannot be made (see Paliya {\em et al.} 2019a, for a detailed discussion). Interestingly, a weak hint for the presence of the soft X-ray excess has also been reported in the bright FSRQ 3C 273 (e.g., Page {\em et al.} 2004; Madsen {\em et al.} 2015), indicating a similar mechanism at work both in \gm-NLSy1 and FSRQs.

\begin{figure*}[t]
\hbox{
\includegraphics[trim = 0 10 0 10, clip, width=\linewidth]{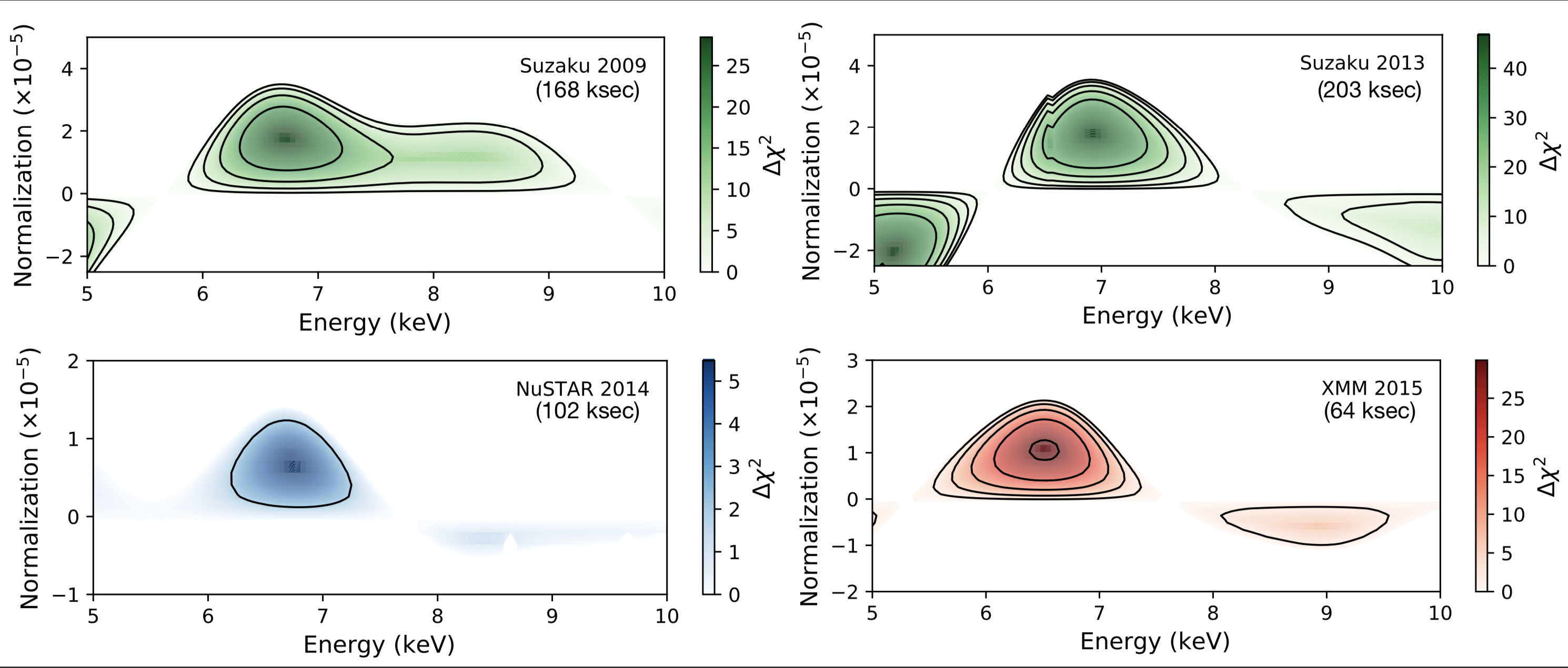}
}
\caption{This diagram illustrates the presence of a weak Fe K$\alpha$ emission line in the X-ray spectrum of 1H 0323+342. To highlight the feature, a Gaussian line (width $\sigma=0.5$ keV) is scanned between 5$-$10 keV energy band for the four different observations, as labeled. Contour lines are at 1-5$\sigma$ confidence level. The data are adopted from Paliya {\em et al.} (2019a).}\label{fig:line}
\end{figure*}

Unlike radio-quiet NLSy1 sources, a strong Fe K$\alpha$ emission line has not been detected in \gm-NLSy1 galaxies. This is likely due to jetted X-ray emission which swamps out any such feature originated from vicinity of the central engine. A tentative evidence of the Fe K$\alpha$ line is found in the deep X-ray observations of one \gm-NLSy1 object, 1H 0323+342 (see Figure~\ref{fig:line}, Abdo {\em et al.} 2009c, Walton {\em et al.} 2013; Kynoch {\em et al.} 2018; Ghosh {\em et al.} 2018). A few researchers have interpreted this weak emission as a relativistic disk line (Walton {\em et al.} 2013; Paliya {\em et al.} 2014) or multiple narrow lines (Kynoch {\em et al.} 2018). By modeling it with a relativistic line profile (Dauser {\em et al.} 2010), a recent study concluded that the inner disk in 1H 0323+342 is probably truncated (Paliya {\em et al.} 2019a). Though this result is tentative, it possibly indicates towards the origin of the jet launching in \gm-NLSy1 galaxies. Lohfink {\em et al.} (2013) studied the radio galaxy 3C 120 and presented a scenario in which inner part of the accretion disk becomes unstable leading to the launch of the jet. A resemblance of the spectral analysis parameters in the case of 1H 0323+342 suggests a similar process at work in the central engine of this \gm-NLSy1 and probably in others as well. 

{\bf Infrared-Ultraviolet:} Many of the \gm-NLSy1 galaxies, e.g., PMN J0948+0022, exhibit a big blue bump at optical-UV energy range which can be interpreted as due to accretion disk emission (Abdo {\em et al.} 2009c; Yao {\em et al.} 2015b; Foschini {\em et al.} 2015). Their synchrotron emission peaks at radio to sub-mm frequencies as revealed by the broadband SED modeling of these sources (discussed later). The IR to UV spectra of other \gm-NLSy1s are found to be dominated by synchrotron radiation (D'Ammando {\em et al.} 2012; Paliya {\em et al.} 2013b, 2019a). Interestingly, even in these sources, the synchrotron peak lies below 10$^{14}$ Hz suggesting them to be LSP type beamed AGNs.

The optical spectrum of a \gm-NLSy1 galaxy consists of broad permitted emission lines of narrow widths. However, the observations also reveal the emission lines to be strong and luminous. The rest-frame equivalent width of the broad lines is found to be greater than 5\AA~for all of the \gm-NLSy1 sources (Oshlack {\em et al.} 2001; Zhou {\em et al.} 2007; Yao {\em et al.} 2015b; Rakshit {\em et al.} 2017). This has two crucial implications: first, all \gm-NLSy1s can be formally classified as FSRQs (e.g., Stickel {\em et al.} 1991; Shaw {\em et al.} 2012) and second, a strong emission line indicates a luminous BLR which in turn suggests an efficient accretion process powering the jets of \gm-NLSy1. In fact, a high-accretion rate has been reported as a defining feature of a NLSy1 galaxy (see, e.g., Boroson 2002; Grupe \& Mathur 2004; Xu {\em et al.} 2012).

{\bf Radio:} As discussed at the beginning of Section~\ref{sec:pres}, a majority of \fermi-LAT detected NLSy1 galaxies exhibit flat radio spectra, though a few compact steep spectrum sources are also found. In particular, one of the \gm-NLSy1s, PKS 2004$-$447, stands out unique as far as the radio properties of this class is concerned. This object exhibits a steep radio spectrum (Oshlack {\em et al.} 2001; Gallo {\em et al.} 2006; Schulz {\em et al.} 2016) suggesting its similarity with the parent population of flat spectrum \gm-NLSy1 galaxies (Berton {\em et al.} 2015). However, its broadband spectral properties resembles well with FSRQs, i.e., beamed AGNs. This includes the detection of a flat X-ray spectrum, a core-dominated one-sided parsec-scale jet, and a Compton dominated broadband SED (Gallo {\em et al.} 2006; Paliya {\em et al.} 2013b). These observations has led to the inclusion of PKS 2004$-$447 in the sample of flat radio spectrum, i.e., beamed, NLSy1 galaxies (e.g., Foschini {\em et al.} 2015).

\begin{figure*}[t]
\hbox{
\includegraphics[scale=0.7]{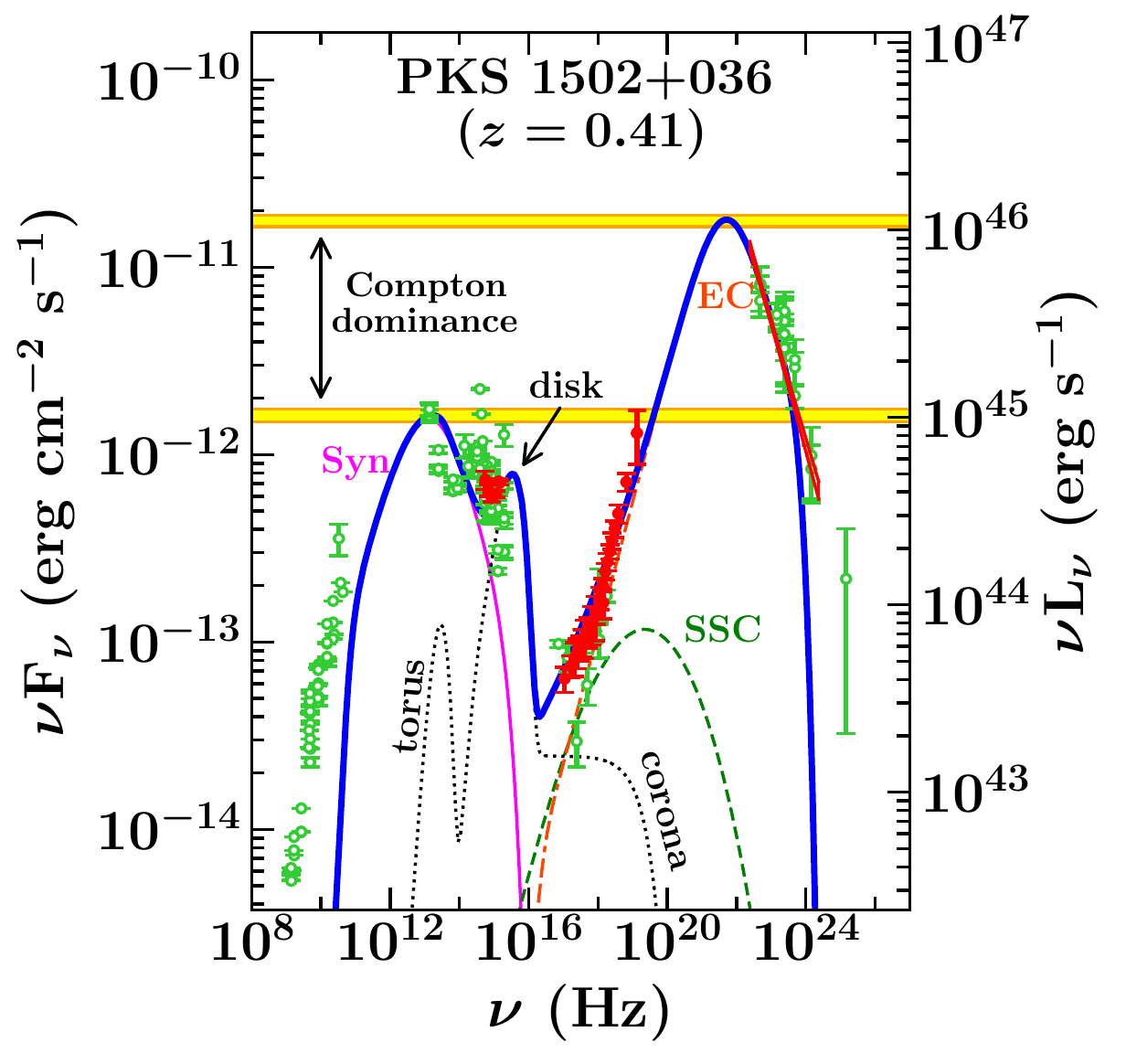}
\includegraphics[scale=0.7]{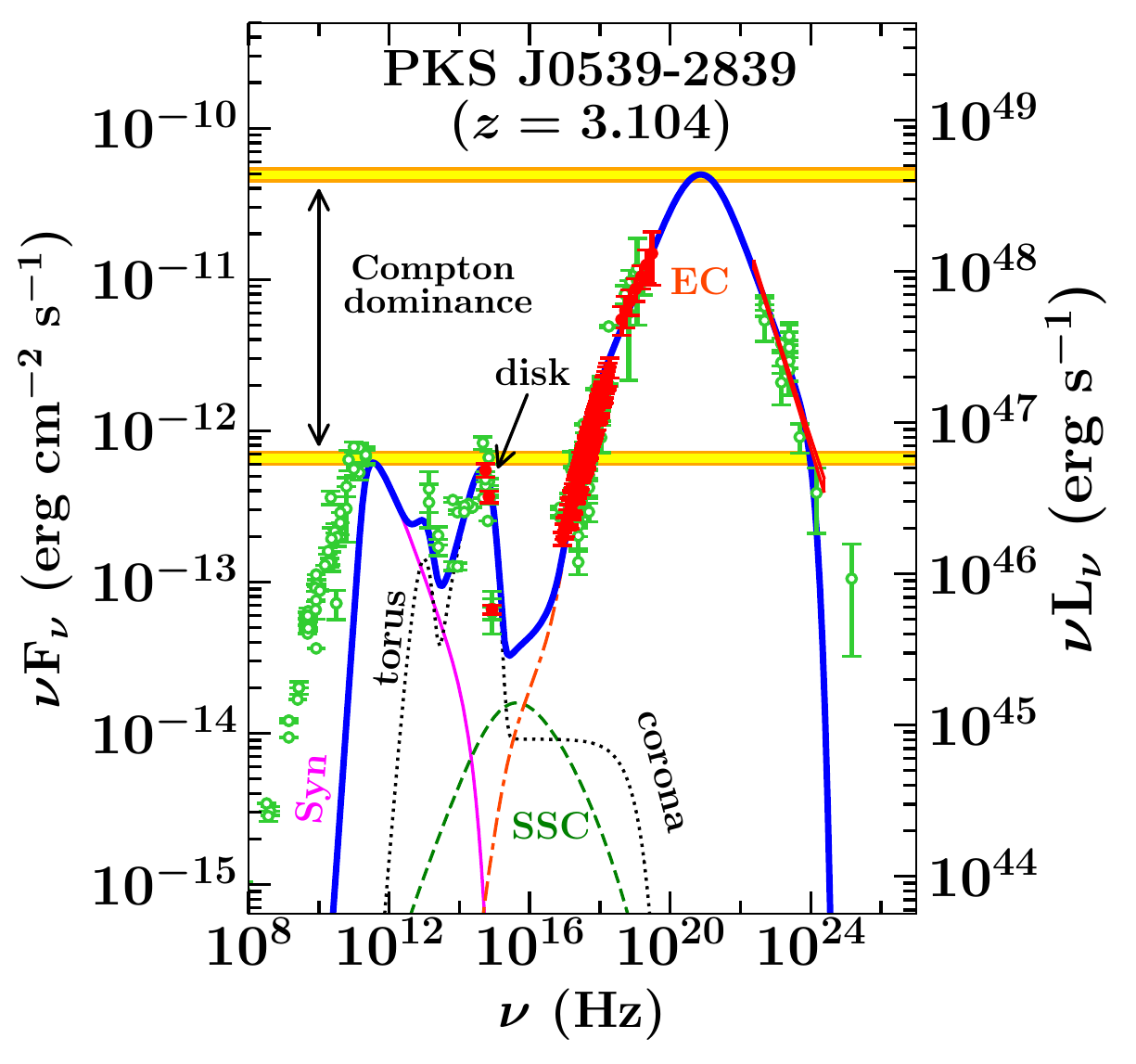}
}
\caption{Left: The multi-wavelength SED of the \gm-NLSy1 galaxy PKS 1502+036. The data used for the modeling is shown with red circles, whereas, archival observations are displayed with green empty circles. Pink thin solid line refers to the synchrotron emission and green dashed and orange dash-dash-dot lines correspond to synchrotron self Compton or SSC and external Compton (EC) processes, respectively. Thermal emission from the dusty torus, accretion disk, and X-ray corona is represented by the black dotted line. Note the location of the synchrotron peak at low (sub-mm) frequencies which leaves the accretion disk radiation naked at optical-UV energies. Another crucial observation is the observed large Compton dominance indicating the prevalence of the external photon field over the magnetic energy density. The broadband SED of other \gm-NLSy1s also exhibit similar features. The data and modeling results are adopted from Paliya {\em et al.} (2019a). Right: The modeled SED of a powerful FSRQ detected with the \fermi-LAT. Other information are same as in the left panel. Note the Compton dominated SED which is a characteristic feature of FSRQs. The data and modeled results are adopted from Paliya {\em et al.} (2019c).}\label{fig:SED}
\end{figure*}

{\bf Broadband SED:} The radio-to-\gm-ray SED of \gm-NLSy1 galaxies show the characteristic double hump structure resembling with that is known for blazars (see Figure~\ref{fig:SED}, Abdo {\em et al.} 2009c; D'Ammando {\em et al.} 2012; Foschini {\em et al.} 2012; Sun {\em et al.} 2015; Yang \& Zhou 2015). As mentioned before, all \gm-NLSy1s are LSP sources and many of them exhibit a pronounced accretion disk emission at optical-UV frequencies. Moreover, the requirement of the external Compton mechanism to explain the \gm-ray emission in these sources provides a supportive evidence about their similarity with FSRQs. The SEDs of the \gm-NLSy1 objects are Compton dominated\footnote{Compton dominance is the ratio of the inverse Compton to synchrotron peak luminosities. A Compton dominated SED is the one in which the inverse Compton radiation dominates the overall multi-wavelength emission, i.e., Compton dominance $>$1. Typically, FSRQs exhibit such behavior (see Figure~\ref{fig:SED}, Finke 2013). On the other hand, BL Lac SEDs are synchrotron dominated (e.g., Yan {\em et al.} 2014) with Compton dominance $<$1.} and a major fraction of their bolometric output radiated in the X-ray to \gm-ray energy range (Figure~\ref{fig:SED}). Since powerful FSRQs are known to exhibit such behavior (cf. Finke 2013; Paliya {\em et al.} 2017a),  it can be concluded that, in a broadband context, both source classes share similar physical properties.

Despite the dearth of multi-wavelength observations, there have been various attempts to characterize the broadband SEDs of the \gm-NLSy1 galaxies and compare them with blazars (see Foschini {\em et al.} 2011; Sun {\em et al.} 2015; Yang \& Zhou 2015; Zhu {\em et al.} 2016; Paliya {\em et al.} 2019a). Particularly, it was noted that these sources host relatively low power jets compared to blazars (cf. Foschini {\em et al.} 2015). Moreover, the Doppler factor, or equivalently bulk Lorentz factor $\Gamma$, of the \gm-NLSy1 jets is reported to be smaller with respect to blazar jets (Paliya {\em et al.} 2019a). Radio observations of these sources also support this finding (Angelakis {\em et al.} 2015; Gu {\em et al.} 2015; Fuhrmann {\em et al.} 2016; Singh \& Chand 2018). A smaller Doppler factor implies a relatively mild boosting which explains the scarcity of the NLSy1 galaxies in the \fermi-LAT catalogs. Furthermore, since the number of the parent population of a beamed AGN is proportional to $\Gamma^2$, a lower Doppler factor also explains an even smaller number of the known parent population members of the \gm-NLSy1 galaxies (Berton {\em et al.} 2015).

\subsection{The Central Engine}
With the settlement of the fact that \gm-NLSy1s host closely aligned relativistic jets similar to blazars, the next major question is to understand the properties of their central engine, i.e., black hole and accretion disk, responsible for the launch of the jet. The most common approach to determine the black hole mass ($M_{\rm BH}$) in AGNs is by using the full width at half maximum (FWHM) of the emission lines and continuum luminosities derived from the single-epoch optical spectroscopy (see, e.g., Vestergaard \& Peterson 2006; Shen {\em et al.} 2011). Interestingly, a comparison of $M_{\rm BH}$ derived from this method for broad-line Seyfert 1s and NLSy1 sources have led to the conclusion that the latter ones host considerably low mass black holes (see, e.g., Grupe \& Mathur 2004; Deo {\em et al.} 2006; Zhou {\em et al.} 2006; Yuan {\em et al.} 2008; Xu {\em et al.} 2012). The presence of narrow optical spectral emission lines indicates a relatively slower virial motion of BLR clouds which, in turn, points towards the low mass of the central gravitating object. A similar conclusion holds when comparing $M_{\rm BH}$ of NLSy1s with blazars since the latter are known to host massive black holes  ($>$10$^8$ \msun, e.g., Shaw {\em et al.} 2012). However, many researchers argued that $M_{\rm BH}$ estimation in NLSy1 (including \gm-ray detected ones) could be biased due to a number of reasons. For example: the observation of narrow emission lines could be due to BLR projection effects under the assumption that the BLR has a disk shaped geometry and NLSy1 galaxies are viewed pole-on (Collin \& Kawaguchi 2004; Jarvis \& McLure 2006; Decarli {\em et al.} 2008). Since NLSy1 objects are highly accreting sources, the negligence of the radiation pressure on BLR clouds caused by the ionizing photons, can also lead to underestimation of $M_{\rm BH}$ (Marconi {\em et al.} 2008).

\begin{figure}[t]
\hbox{
\includegraphics[width=\linewidth]{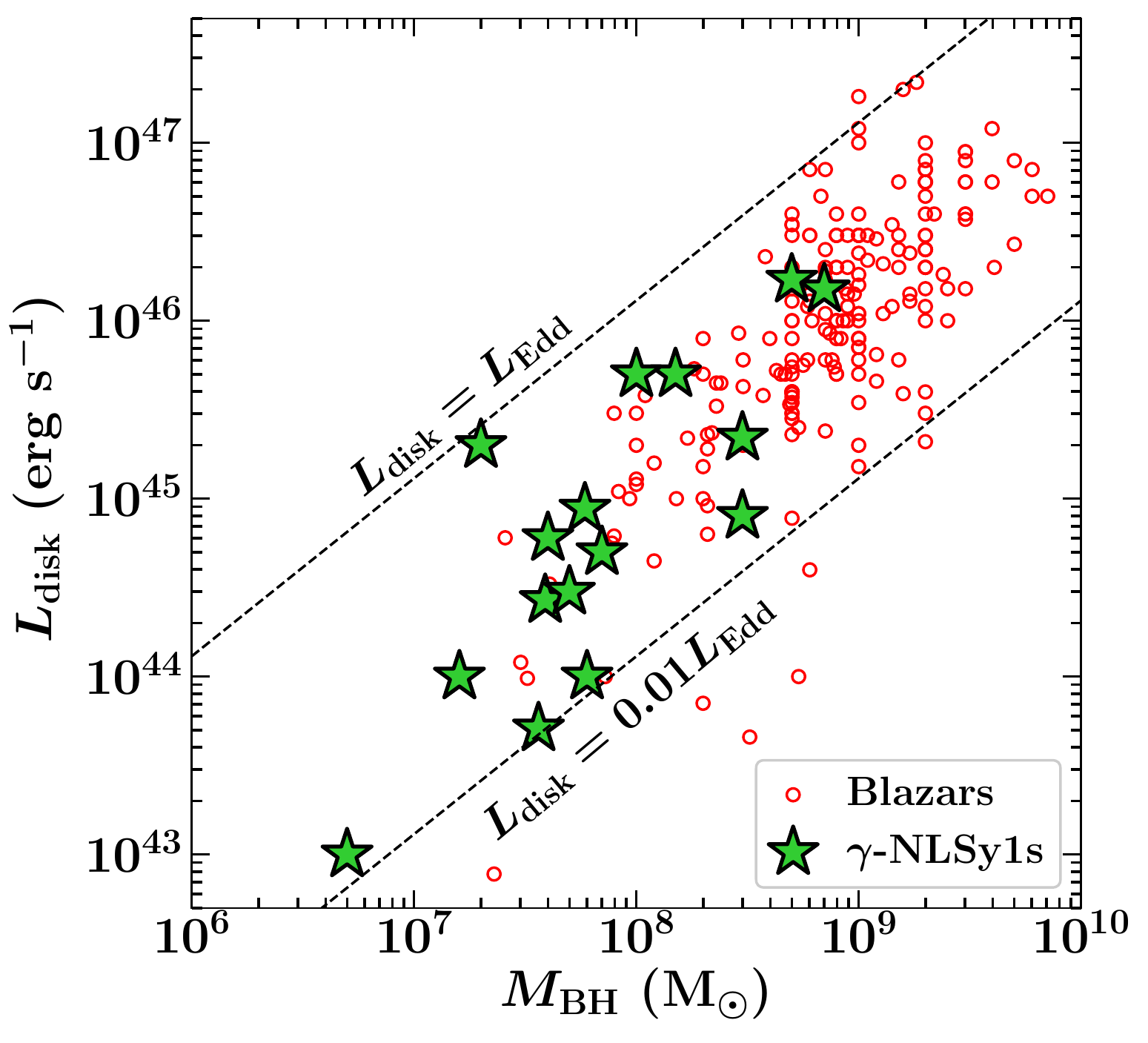}
}
\caption{Properties of the central engine in \gm-NLSy1 galaxies. As can be seen, a majority of them occupy the low disk luminosity-low black hole mass regime with respect to blazar population. However, note that all of them accrete at $>$1\% of the Eddington luminosity. The plot is adopted from Paliya {\em et al.} (2019a).}\label{fig:engine}
\end{figure}

There are many other observations which disagree with higher $M_{\rm BH}$ in NLSy1 galaxies as predicted by above mentioned hypotheses. If all NLSy1s are viewed pole-on, then we should expect a larger or at least comparable radio-loud fraction and variability for NLSy1 galaxies with respect to their broad-line counterparts due to beaming effect. These predictions are not confirmed (Rakshit \& Stalin 2017). Moreover, the Lorentzian line profile of NLSy1s and other population A sources of the EV1 (Sulentic, Marziani, \& Dultzin-Hacyan 2000, Sulentic \& Marziani 2015) suggests that the geometry of the BLR is not flattened (see Kollatschny \& Zetzl 2011, 2013). A similar findings has also been reported when comparing optical spectra with X-ray properties of NLSy1s (Vietri {\em et al.} 2018). A few other not-so-common methods, independent of the orientation effects, have also been adopted, e.g., spectropolarimetry (Baldi {\em et al.} 2016), the excess variance determined from the X-ray variability (Pan {\em et al.} 2018), to estimate $M_{\rm BH}$ of the individual \gm-NLSy1 objects and generalizing these techniques to the population is yet to be done.

Another method to determine $M_{\rm BH}$ in AGNs is by reproducing the optical-UV bump with an accretion disk model (e.g., Calderone {\em et al.} 2013, Ghisellini {\em et al.} 2014, but see Calderone, D'Ammando, \& Sbarrato 2018, for recent updates). Usually, the results derived from this method are found be in agreement with that obtained from the single-epoch optical spectroscopy (see Ghisellini \& Tavecchio 2015). By applying this technique to \gm-NLSy1s, it was recently shown that they also follow the same trend within the uncertainties associated with the virial method (Paliya {\em et al.} 2019a). The derived results confirm that \gm-NLSy1 galaxies, on average, host low mass black holes with respect to blazar population. Moreover, their accretion disk luminosities are also on the lower side. This is demonstrated in Figure~\ref{fig:engine}. Interestingly, it can be seen that all \gm-NLSy1 galaxies have accretion rate $>$1\% in Eddington units, i.e., they are efficient accretors. Therefore, though the central engines of \gm-NLSy1s are less powerful, they are as efficient as in FSRQs.

\subsection{Accretion-Jet Connection}
It has been postulated a long ago that jets and accretion process are significantly connected (Rawlings \& Saunders 1991). If jets are powered by accretion, then a positive correlation between the accretion disk luminosity and the jet power is expected. Indeed, blazars are found to follow such a trend (Celotti \& Ghisellini 2008; Ghisellini {\em et al.} 2014; Inoue {\em et al.} 2017; Paliya {\em et al.} 2017a; Chen 2018). Extending this hypothesis to the realm of the \gm-NLSy1 galaxies, even though the number of known sources is small, allows one to explore the jet-disk connection in a different AGN environment.

\begin{figure*}[t]
\hbox{
\includegraphics[scale=0.57]{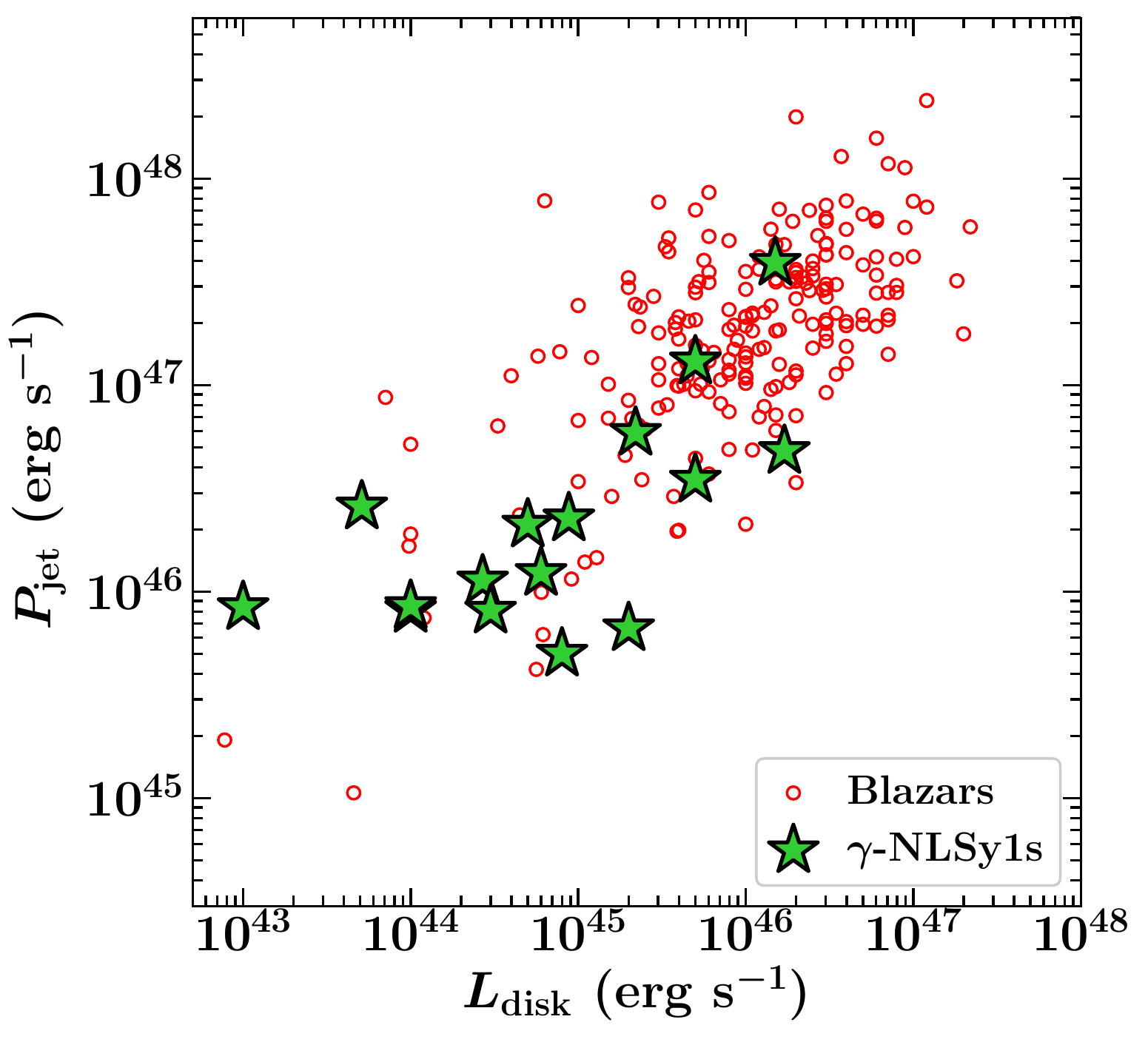}
\includegraphics[scale=0.57]{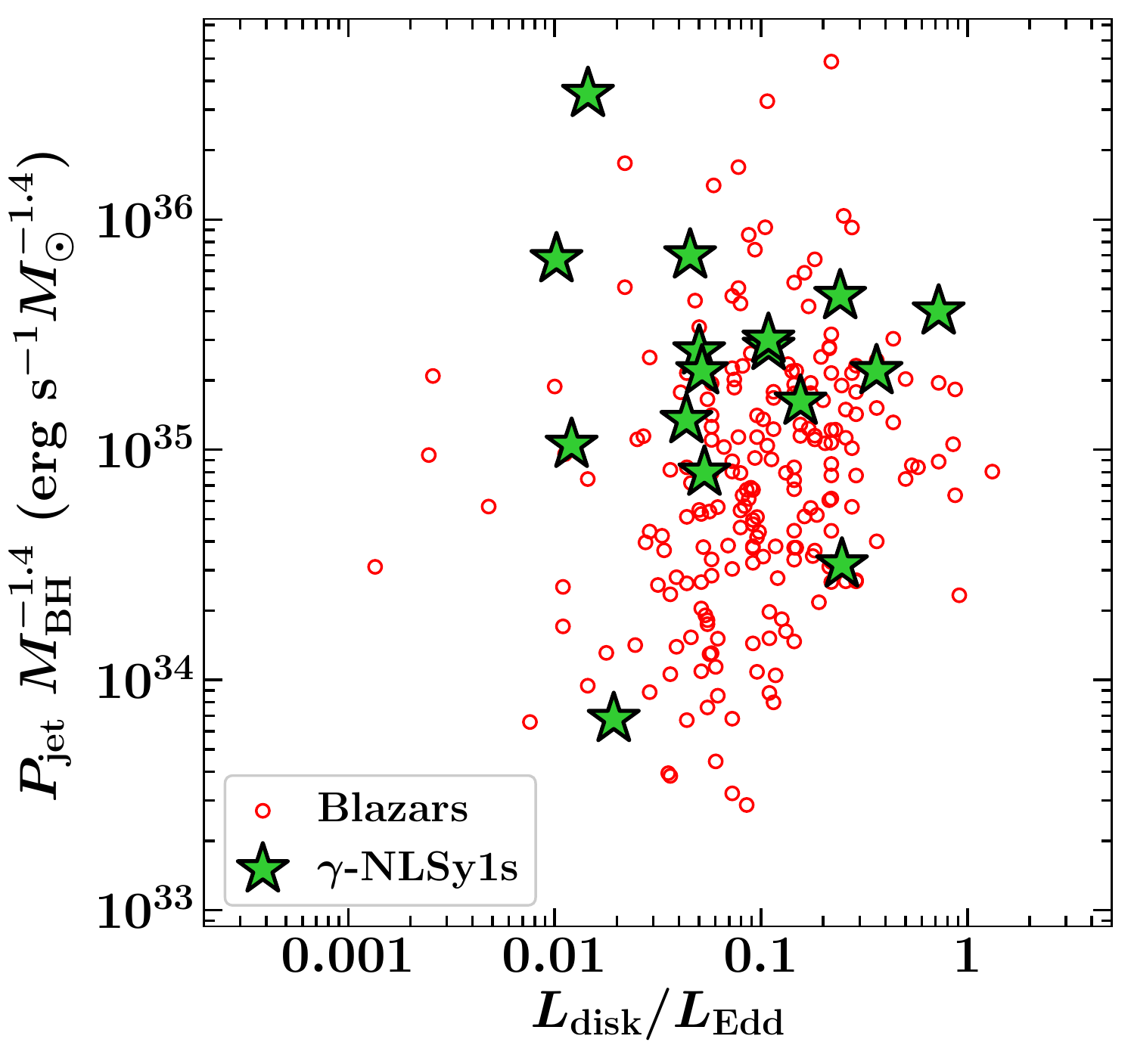}
}
\caption{Left: The jet power ($P_{\rm jet}$) as a function of the accretion disk luminosity derived from the leptonic jet modeling. As can be seen, a majority of the \gm-NLSy1 galaxies lie at the low-end of the observed correlation. Note the blazar sample mainly consists of FSRQs (Paliya {\em et al.} 2017a). This is because the presence of broad optical emission lines enables a far accurate determination of the accretion power in FSRQs compared to BL Lac objects. Right: Same as left, but after normalizing the parameters for the central black hole mass. The diagrams are adopted from Paliya {\em et al.} (2019a).}\label{fig:jet}
\end{figure*}

Various recent studies have revealed that the \gm-NLSy1 sources lie at the low-end of the disk-jet connection observed in blazars (see Figure~\ref{fig:jet}, left panel, Foschini {\em et al.} 2015; Paliya {\em et al.} 2019a) and therefore they can be considered as the low-luminosity counterparts of powerful blazars, mainly FSRQs. Furthermore, a comparison of the accretion disk luminosity and the jet power, after correcting for the central black hole mass (e.g., following Heinz \& Sunyaev 2003), has led to the conclusion that \gm-NLSy1 jets are as powerful as in FSRQs (see Figure~\ref{fig:jet}, right panel; Foschini 2012). This observation suggests that since NLSy1 galaxies are posed as young AGNs in the evolution scheme, \gm-NLSy1s can be thought as young, low power, precursor of FSRQs. These source also well-fit in the proposed jet evolution scenario in which FSRQs evolve to low-luminosity BL Lac objects as the accretion power decreases (B{\"o}ttcher \& Dermer 2002; Cavaliere \& D'Elia 2002). Accordingly, the sequence of jet evolution can be as follows: \gm-NLSy1$\rightarrow$FSRQs$\rightarrow$BL Lacs (see also Foschini 2017).

\subsection{Jets at Kiloparsec and Parsec Scales and Kinematics}
Most of the radio-loud NLSy1 galaxies were found to exhibit compact ($\lesssim$300 pc) radio structures at GHz frequencies (Ulvestad {\em et al.} 1995; Zhou {\em et al.} 2006; Whalen {\em et al.} 2006; Yuan {\em et al.} 2008). This was mainly due to limited spatial resolution of FIRST survey  (5$^{\prime\prime}$, Helfand {\em et al.} 2015) which can resolve only structures larger than 10 kpc at $z=0.8$, the typical redshift up to which NLSy1s are known. As of now, only about a dozen NLSy1 galaxies are known hosting kiloparsec or kpc scale radio jets (Doi {\em et al.} 2012; Richards \& Lister 2015; Congiu {\em et al.} 2017, Berton {\em et al.} 2018; Rakshit {\em et al.} 2018; Singh \& Chand 2018). The \gm-NLSy1 galaxies are compact at kpc scale (cf. Giroletti {\em et al.} 2011; D'Ammando {\em et al.} 2012; Orienti {\em et al.} 2015) and only 1H 0323+342 exhibit $>$10 kpc jet (e.g., Ant{\'o}n {\em et al.} 2008). On the other hand, many of them are resolved at parsec-scale, often showing milli-arcsec core-jet structure, when observed with Very Long Baseline Interferometry (see, e.g., Wajima {\em et al.} 2014, Doi {\em et al.} 2015, Berton {\em et al.} 2018).

\begin{figure}[t]
\hbox{
\includegraphics[width=\linewidth]{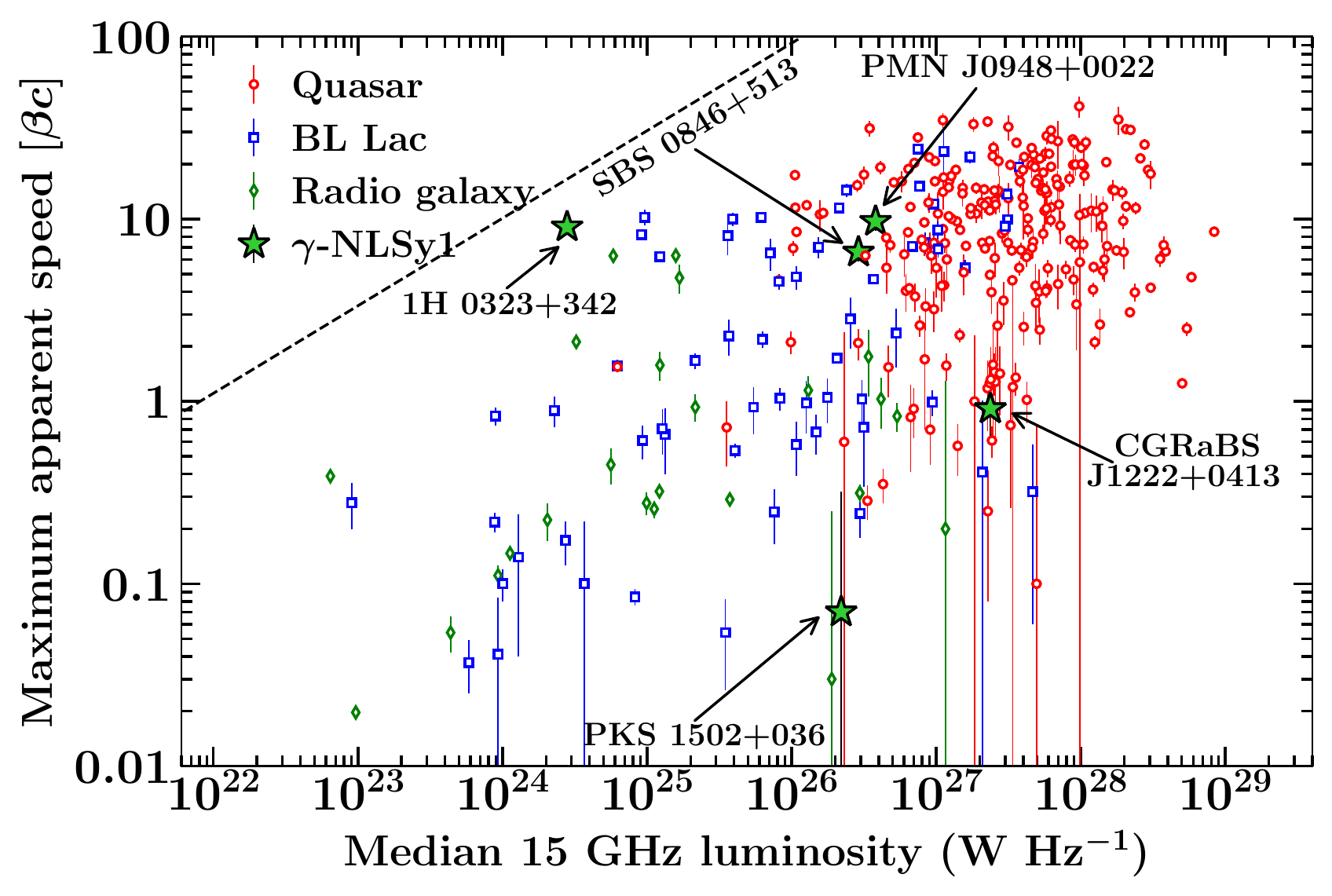}
}
\caption{Variation of the maximum jet speed as a function of the 15 GHz radio luminosity for a sample of jetted AGNs using VLBA observations under MOJAVE program. The \gm-NLSy1 galaxies are labeled. The dashed line represents the upper limit of the survey. Plotted data are taken from Lister {\em et al.} (2019).}\label{fig:kinematics}
\end{figure}

The parsec-scale jet kinematics study of about half-a-dozen \gm-NLSy1 galaxies at 15 GHz has been carried out using Monitoring Of Jets in Active galactic nuclei with VLBA Experiments (MOJAVE) program (Lister {\em et al.} 2009, 2016, 2019). Superluminal motion has been observed from three \gm-NLSy1 galaxies (Figure~\ref{fig:kinematics}): 1H 0323+342 ($\beta_{\rm app,~c}=9.1\pm0.3$, Lister {\em et al.} 2016), SBS 0846+513 ($\beta_{\rm app,~c} =6.6\pm0.8$, Lister {\em et al.} 2019), PMN J0948+0022 ($\beta_{\rm app,~c} =9.1\pm0.3$, Lister {\em et al.} 2019). The bulk Lorentz factor ($\Gamma>v_{\rm app}/c$) and the jet viewing angle ($\theta_{\rm v}<2\arctan(c/v_{\rm app})$) can be constrained from the observed superluminal motion. The derived parameters suggest the presence of beamed jet emission in \gm-NLSy1 galaxies (see also Lister 2018, for a recent review).

According to the dynamical evolution models, compact parsec-scale radio jets turn into kpc-scale ones which ultimately evolve to large jet-lobe radio structures of the size of $\sim$hundreds of kpc (Readhead {\em et al.} 1996). Keeping in mind the fact that \gm-NLSy1 galaxies are likely to be young systems, existing observations provide supportive evidences that radio jets in these sources are probably in their early phase of the evolution. Since NLSy1 galaxies are expected to have dense interstellar medium (ISM) due to the observation of enhanced star-formation activities (Caccianiga {\em et al.} 2015), another possibility could be that the jet may remain confined within the nuclear region of the host galaxy (Marecki {\em et al.} 2003; Kawakatu {\em et al.} 2008). However, this explanation may not hold for the jets in \gm-NLSy1 galaxies due to their peculiar orientation with respect to our line of sight and the observation of variable \gm-ray emission which suggests that the jet is able to pierce the ISM.

\begin{figure*}[t]
\hbox{
\includegraphics[trim= 100 80 100 80, clip, scale=0.315]{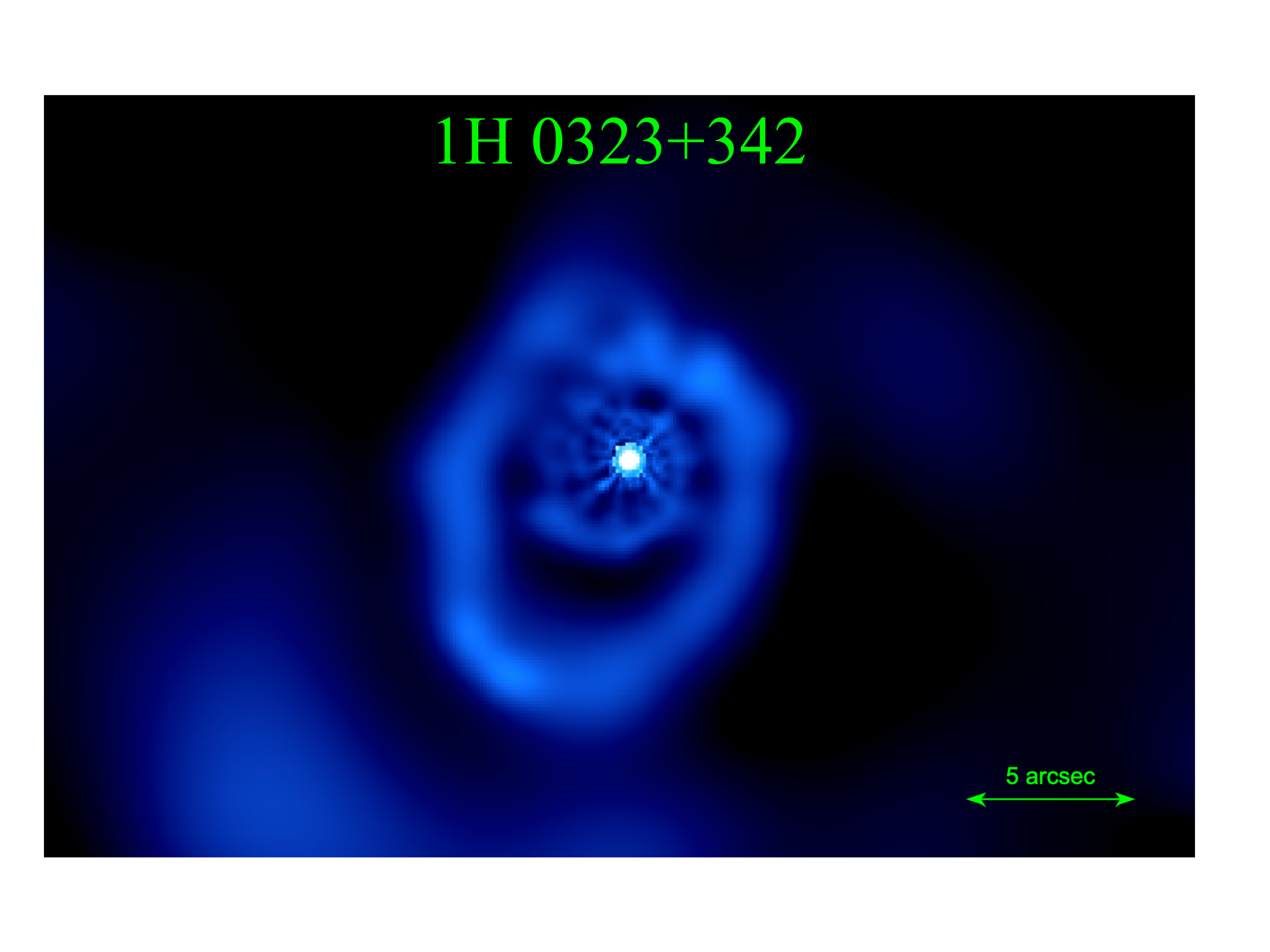}
\includegraphics[trim= 170 80 80 70, clip, scale=0.31]{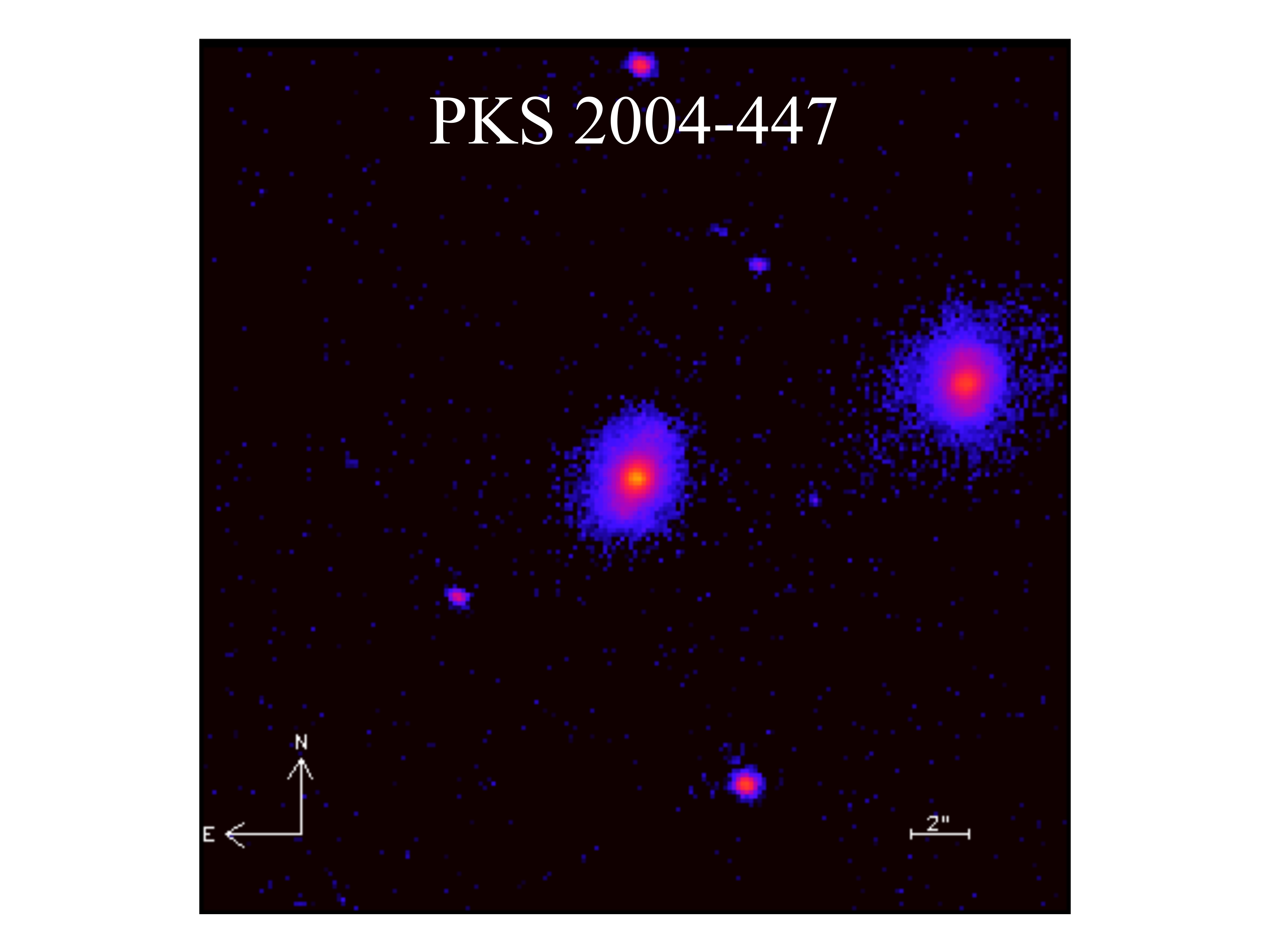}
}
\hbox{
\includegraphics[trim= 200 200 200 180, clip, scale=0.465]{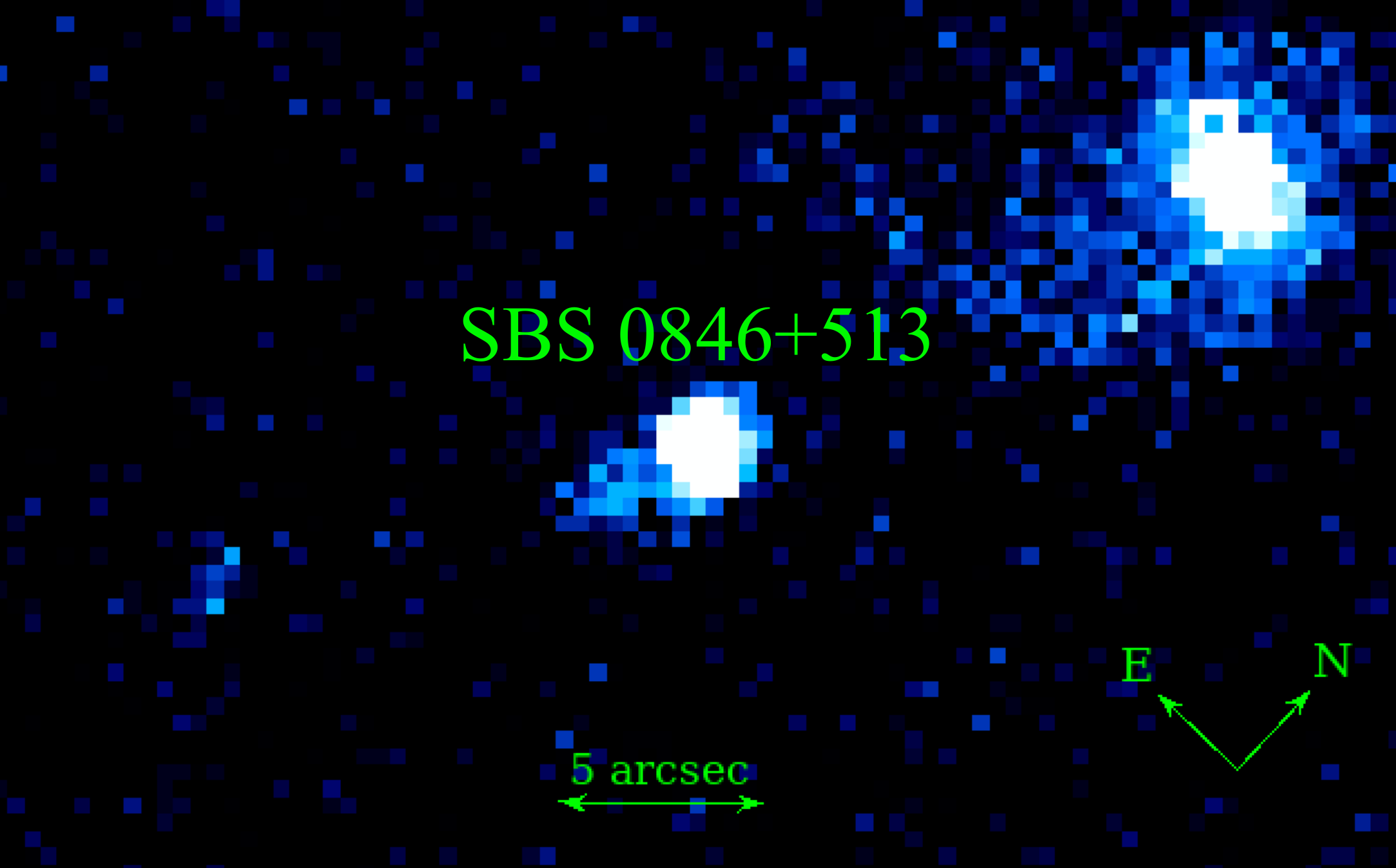}
\includegraphics[trim= 200 100 200 120, clip, scale=0.33]{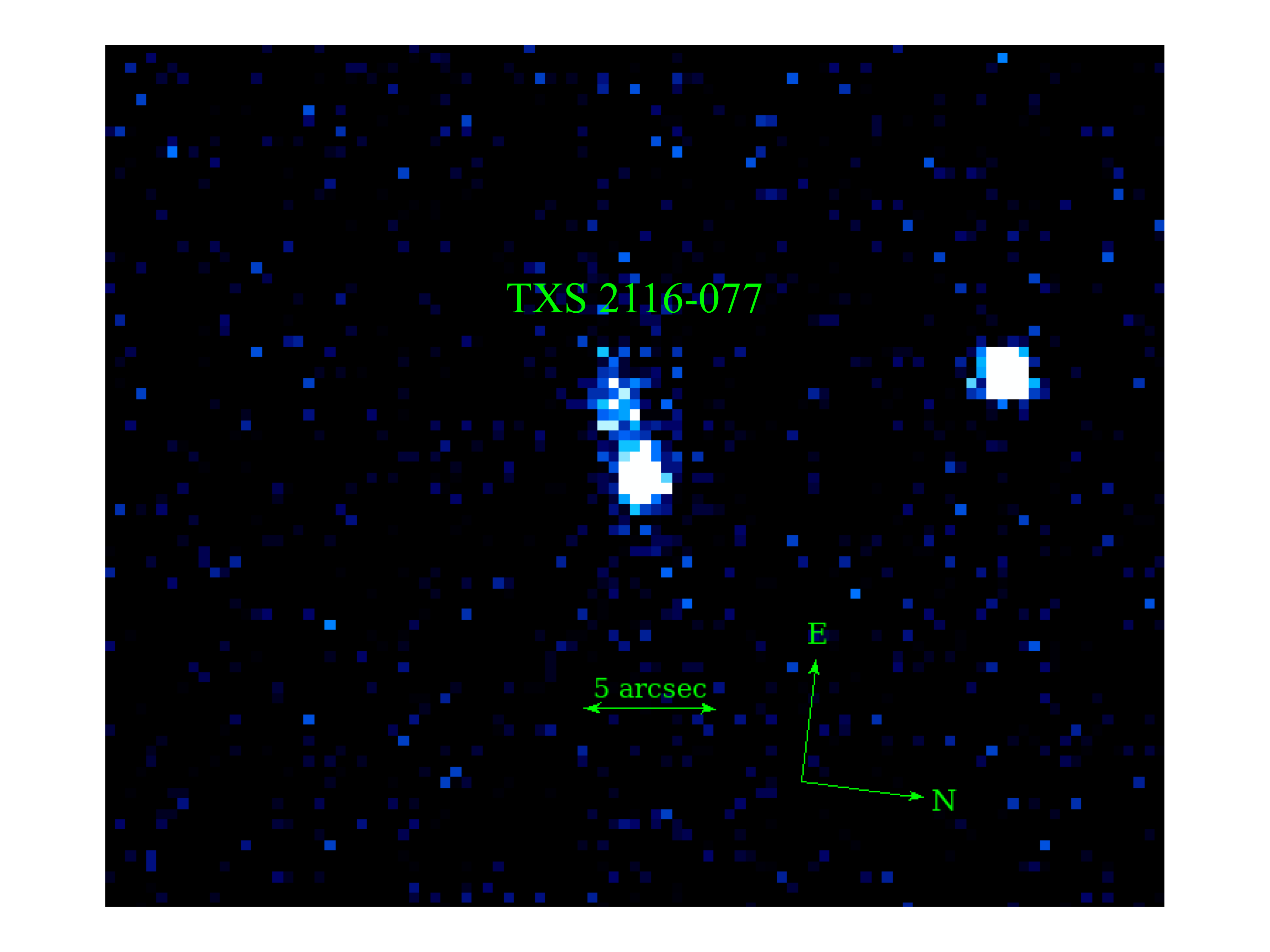}
}
\caption{Left: Host galaxy morphology of \gm-NLSy1 galaxies. Top left: adaptively smoothed {\it Hubble Space Telescope}/WFPC2 image of 1H 0323+342 with the filter F702W (6900\AA). Adapted from Foschini (2011). Top right: $J$-band image of PKS 2004$-$447 obtained from Very Large Telescope at European Southern Observatory in Paranel, Chile. Adopted from Kotilainen {\em et al.} (2016) and used with permission from the AAS. Bottom: Low resolution SDSS $z$-filter images of \gm-NLSy1 galaxies SBS 0846+513 (left) and TXS 2116$-$077 (right).  Note the traces of extended emission in both sources likely due to ongoing or recent mergers. PKS 2004$-$447 is claimed to be evolving via secular processes (Kotilainen {\em et al.} 2016).}\label{fig:host}
\end{figure*}

\subsection{Host Galaxy and Large-Scale Environment}
The evolution of a galaxy is very much affected by the central AGN it hosts and vice versa (e.g., Fabian 2012; King \& Pounds 2015). The host galaxy controls the gas supply to the central black hole and thus directly affecting its activity level. On the other hand, AGN regulates the evolution of the host via various feedback processes. Furthermore, the host galaxy morphology, e.g., elliptical vs. disk/late-type, also enables us to understand the modes of galaxy evolution, e.g., secular versus merger driven. Concerning the jetted AGNs, recent host galaxy imaging studies focusing on Type 2 radio-loud and radio-quiet quasars have revealed the mergers and disturbed morphologies to be more frequently associated with the former (Ramos Almeida {\em et al.} 2012, 2013; Chiaberge {\em et al.} 2015), suggesting a close connection of triggering of jets with the galaxy mergers. Also, radio-loud quasars, including blazars, are known to be hosted in massive elliptical galaxies (Dunlop {\em et al.} 2003; Olgu{\'{\i}}n-Iglesias {\em et al.} 2016) which are proposed to form via mergers (Toomre \& Toomre 1972).

The host galaxy morphology studies of mostly radio-quiet NLSy1 galaxies have revealed that a major fraction of them reside in late-type disk or spiral galaxies with pseudo bulges and stellar bars (Crenshaw {\em et al.} 2003, Ohta {\em et al.} 2007, Orban de Xivry {\em et al.} 2011, Mathur {\em et al.} 2012, J{\"a}rvel{\"a} {\em et al.} 2018). NLSy1s hosted in spirals show a higher fraction of nuclear dust arms and bars compared to their broad line counterparts (Crenshaw {\em et al.} 2003; Deo {\em et al.} 2006). Moreover, radio-quiet NLSy1 also do not show strong evidence for mergers (Krongold {\em et al.} 2001). Overall, these observations suggest the growth of the central black hole in radio-quiet NLSy1 galaxies to be primarily driven by secular processes rather than by merger-induced accretion. 

The host galaxy imaging of \gm-NLSy1 galaxies is recently done for a handful sources. The {\it Hubble Space Telescope} imaging of the nearest \gm-NLSy1, 1H 0323+342, reveals a clear extended structure, possibly a spiral arm or a star-forming ring due to a recent merger (see Figure~\ref{fig:host}; Zhou {\em et al.} 2007; Ant{\'o}n {\em et al.} 2008; Le{\'o}n Tavares {\em et al.} 2014). Similarly, a few other \gm-NLSy1 galaxies, e.g., SBS 0846+513 and TXS 2116$-$077, exhibit either an ongoing merger or traces of violent activities in the past (see Figure~\ref{fig:host}, Paliya {\em et al.} 2018; Yang {\em et al.} 2018). A systematic study of the host galaxy morphology of a large sample ($>$20) of radio-loud NLSy1 galaxies, including \gm-NLSy1, also indicate a major fraction ($\sim$70\%) of them to participate in merger or have a close companion\footnote{https://zenodo.org/record/1406036} (see also Berton {\em et al.} 2019). A few other \gm-NLSy1 galaxies do not show evidence of mergers and rather support secular driven black hole growth and jet launching (e.g., Kotilainen {\em et al.} 2016; Olgu{\'{\i}}n-Iglesias {\em et al.} 2017). For a few sources, claims have also been made them to reside in elliptical galaxies similar to blazars (D'Ammando {\em et al.} 2017, 2018). However, a majority of these sources are found to be hosted in young late-type galaxies with pseudo bulges and presence of disk and bars are also reported.

The average large-scale environment density for NLSy1 galaxies is found to be lesser compared to the broad-line Seyfert 1s and jetted AGNs. This observation provide another confirmation about the young nature of NLSy1s. This is because galaxy-galaxy interactions are more frequent in denser regions leading to evolution of the galaxy morphology to ellipticals, i.e., early-type, though the speed of transformation also depends on the environment density (Dressler 1980; Park \& Choi 2009; Chen {\em et al.} 2017). However, among NLSy1 sources, radio-loud ones tend to reside preferentially in denser surroundings (J{\"a}rvel{\"a} {\em et al.} 2017) suggesting a close connection between the large-scale environment  and radio-loudness similar to that found for jetted AGNs.

\section{Future: Outstanding Questions}\label{sec:fut}
The broadband campaigns organized to study various aspects of the \gm-NLSy1 galaxies have significantly improved our current understanding of relativistic jets, their origin and the connection between the \gm-NLSy1s and blazars. However, these discoveries have also raised even more outstanding questions. A brief description about what more can be done with currently working observational facilities and what to expect from the next generation missions is provided below.

{\bf New \gm-NLSy1 galaxies:} In the first $\sim$10 years of \fermi-LAT operation, less than two dozen radio-loud NLSy1 galaxies are identified as \gm-ray emitters. In contrast, more than 2500 \gm-ray blazars are listed in the recent released fourth catalog of \fermi-LAT detected AGNs (The Fermi-LAT collaboration 2019). Though all sky survey by the \fermi-LAT is continued implying an even deeper exposure of the \gm-ray sky, one needs to look for alternative strategies to identify more \gm-ray emitting NLSy1s, other than time-averaged analyses used in \fermi-LAT catalogs and individual studies (e.g., Paliya {\em et al.} 2018). Two possible approaches that one can adapt are as follows:
\begin{enumerate}
\item Search for \gm-ray transient NLSy1 galaxies.
\item Explore the optical spectroscopic properties of the known \gm-ray emitting blazars.
\end{enumerate}

The first method considers the fact that there could be a few radio-loud NLSy1 galaxies that might have exhibited low-amplitude \gm-ray flares since the launch of \fermi~satellite. When combining long time ($\sim$8-10 years) \fermi-LAT data, such events can easily get averaged out leading to the non-detection of the source. However, by generating \gm-ray light curves (weekly-to-monthly timescales), one can look for short-timescale flaring episodes. If detected, this may provide a conclusive evidence supporting the presence of the beamed emission. This exercise has been successfully applied to search \gm-ray emission from high-redshift ($z>2$) blazars\footnote{https://agenda.infn.it/event/15975/contributions/32078/} which are not present in any \fermi-LAT catalogs (e.g., Liao {\em et al.} 2018). This exercise can also be extended to non-radio detected NLSy1 galaxies as a few of them may harbor relativistic jets (see, e.g., L{\"a}hteenm{\"a}ki {\em et al.} 2018).

A radio-loud quasar is classified as a blazar based on its broadband physical properties, such as flat radio spectrum, high brightness temperature, and variability. From the optical spectroscopic perspective, the only criteria is the rest-frame line EW, $>$5\AA~for FSRQs. On the other hand, NLSy1 galaxies are solely identified based on their optical spectral characteristics as mentioned in Section~\ref{sec:intro} Therefore, it is advisable to explore the optical spectra of the known \gm-ray blazars, particularly FSRQs, and determine if some of them turn out to be genuine NLSy1s. Recently, a few studies have adopted this approach and successfully identified a few FSRQs as \gm-NLSy1s  (Yao {\em et al.} 2015b, 2019; Paliya {\em et al.} 2018). Additionally, more than 30\% of \fermi-LAT detected AGNs are blazars of uncertain type or unidentified \gm-ray sources, i.e. sources with no known counterparts (The Fermi- LAT collaboration 2019). A careful investigation of the optical spectral properties of these objects could lead to identification of new \gm-NLSy1 galaxies as recently shown by Paiano {\em et al.} (2019). The fact that so far no \gm-ray space mission is planned after \fermi, this is probably the best strategy to increase the sample size of the known \gm-NLSy1 sources.

{\bf Disk-corona-jet connection:} Among all of the known \gm-NLSy1 objects, the most promising candidate to reveals the disk-corona-jet connection is 1H~0323+342 (see Paliya {\em et al.} 2014; Yao {\em et al.} 2015a; Landt {\em et al.} 2017; Ghosh {\em et al.} 2018). This is the only \gm-NLSy1 source which has shown evidences, albeit low-significance, for the presence of Fe K$\alpha$ line in its soft X-ray spectrum. Moreover, its 0.3$-$10 keV emission is found to be dominated by the coronal emission, except during elevated \gm-ray activity states when jet emission takes over. In order to reveal the underlying features associated with the disk-corona interaction, a carefully planned campaign is needed in which the source can be regularly monitored with \swift-XRT (one 1 ksec pointing per week is sufficient) to keep an eye on its activity state\footnote{The sky surveying capability of \fermi-LAT also enables a continuous monitoring of the jet activity of 1H 0323+342 or any other source, however, being faint in \gm-rays, there could be ambiguity in determining the level of the jet activity (see Figure~\ref{fig:1H323_g_lc}).}. At the time of the lowest flux state (with respect to historical observations, see Figure~\ref{fig:1H323_g_lc}), deep X-ray observations from \xmm~and \nustar~can be triggered to explore the coronal features when the jet activity is the lowest. Depending on the availability of the telescope time, a similar approach can be adopted for other \gm-NLSy1 galaxies, especially those which are faint indicating a weak jet. Identifying new \gm-NLSy1s will also enable us to determine the most promising candidates for a deep X-ray spectral study. Looking at the future, upcoming missions like {\it Athena} and {\it XRISM} will further probe the intriguing nature of \gm-NLSy1 galaxies in the X-ray band (see, e.g., Cappi {\em et al.} 2013).

{\bf Host galaxy imaging:} The fact that \gm-NLSy1 galaxies have weak jets compared to blazar population, makes them excellent candidates to probe the host galaxy environment. Using existing observations, e.g., with SDSS, the most promising sources can be selected for deep NIR imaging with big (8 m or 10 m class) telescopes. This needs to be done for a carefully selected sample of NLSy1 galaxies (both radio-loud and radio-quiet, see, e.g., J{\"a}rvel{\"a} {\em et al.} 2018) to make a meaningful comparison of their host morphologies. Moreover, many \gm-NLSy1 galaxies exhibit traces of mergers, e.g., SBS 0846+513 (Figure~\ref{fig:host}), thus it is very important to carry out integral field spectroscopic observations of the sources with disturbed morphologies to determine the role of mergers in triggering jets. Followup observations from Atacama Large Milimeter/submilimeter Array (ALMA) will allow us to procure high-resolution, large-scale, images of the molecular gas in host galaxies. Since mergers are also known to trigger starburst activities (Di Matteo {\em et al.} 2005; Luo {\em et al.} 2014), ALMA observations can be used to explore the dynamics of the molecular component which will throw more light on the mechanisms of starbursts-AGNs, associated feedback processes, such as outflows of molecular gas, bubbles and winds, and most importantly how these physical mechanisms interact with the relativistic jet. Furthermore, observations from the next generation giant observatories, such as Thirty Meter Telescope and European Extremely Large Telescope, equipped with integral field unit will be crucial to unravel the jet triggering mechanism in \gm-NLSy1 galaxies.

\begin{figure}[t]
\hbox{
\includegraphics[width=\linewidth]{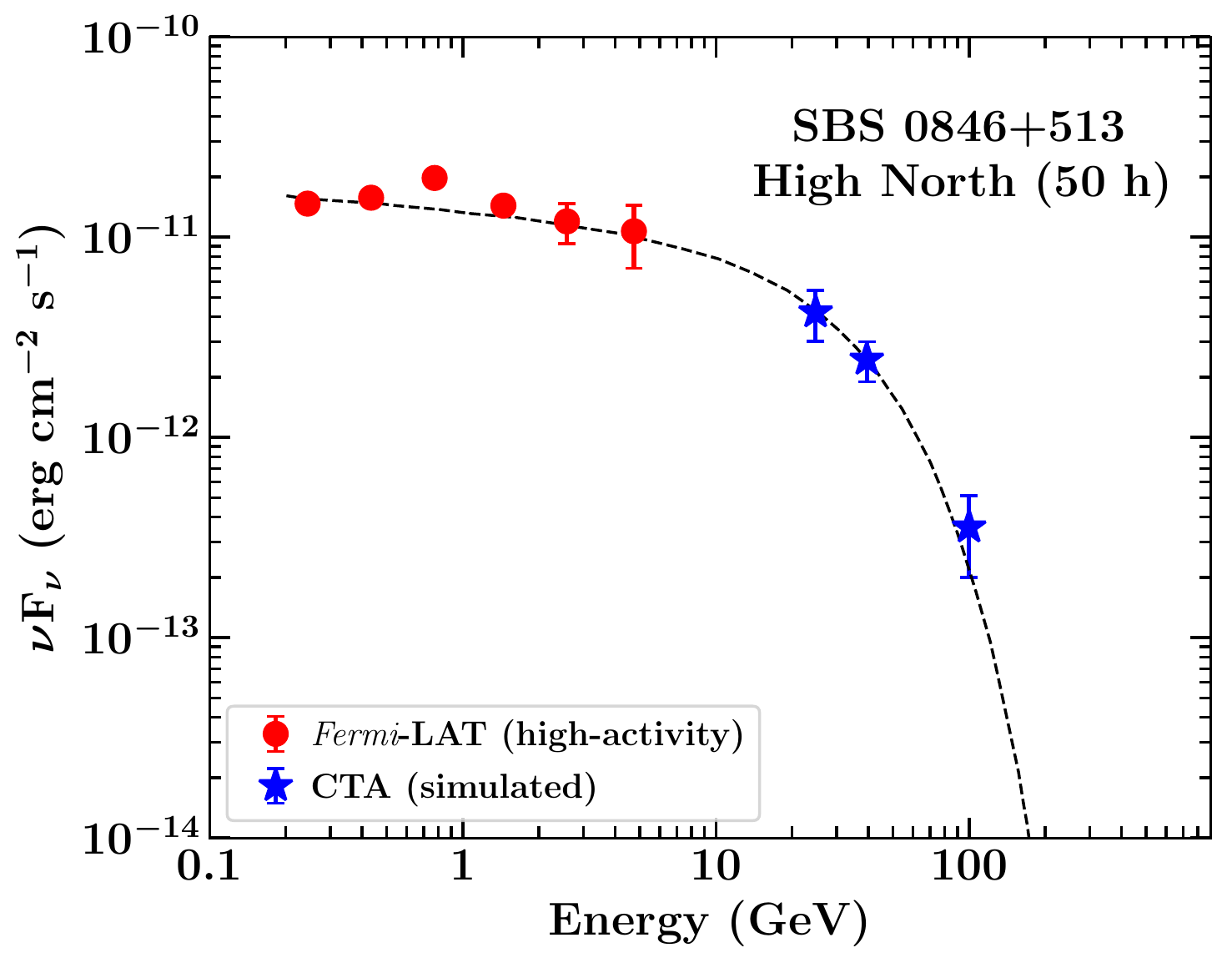}
}
\caption{Gamma-ray SED of the \gm-NLSy1 galaxy SBS 0846+513 representing a high jet activity state. Red data points refer to the 2013 April-July \gm-ray flare of the source, whereas, blue stars correspond to the simulated fluxes for 50 hr of CTA-North exposure. The \fermi-LAT and CTA data are adopted from Paliya {\em et al.} (2016) and Romano {\em et al.} (2018), respectively. The black dashed line represents the input model used for the simulation whose details can be found at Romano {\em et al.} (2018).}\label{fig:cta}
\end{figure}

{\bf NLSy1 galaxies with Cherenkov Telescope Array (CTA):} None of the \gm-NLSy1 galaxies are detected with currently operating atmospheric Cherenkov telescopes, e.g. VERITAS and MAGIC. Since these objects exhibit soft \gm-ray spectrum similar to FSRQs, their detection probability with the very high energy (VHE, $>$100 GeV) facility remains low. However, the fact that blazars often show a harder \gm-ray spectrum during elevated activity state (see, e.g., Paliya {\em et al.} 2019b), makes them a potential candidate for VHE detection since \gm-NLSy1 galaxies are expected follow a similar trend. Indeed, keeping in mind the unprecedented sensitivity of the upcoming VHE observatory, CTA, Romano {\em et al.} (2018) performed a series of dedicated simulations taking into account the spectral parameters and attenuation by the extragalactic background light and found that prospects of VHE detection of \gm-NLSy1s are promising. Particularly, SBS 0846+513, PMN J0948+0022, and PKS 1502+036 are likely to get detected by CTA during their high jet activity state (see Figure~\ref{fig:cta}). These observations will provide conclusive evidences about the physical conditions and emission mechanisms of \gm-NLSy1 jets.

\section{Conclusion}\label{sec:con}
The main observational findings revealing the nature of \gm-NLSy1 galaxies can be summarized as follows:
\begin{enumerate}
\item The \gm-NLSy1 sources, similar to more-common NLSy1s, are young, highly accreting AGNs hosting low mass black holes and associated with late-type galaxies having pseudo bulges and bars.
\item There are evidences indicating \gm-NLSy1s, including radio-loud NLSy1, to reside in denser large-scale environments where galaxy-galaxy interactions are frequent. In fact, many \gm-NLSy1s exhibit traces of past or ongoing mergers.
\item The \gm-NLSy1 galaxies host comparatively small-scale, low-power jets as confirmed from the deep radio imaging and broadband SED modeling. These observations indicate the \gm-NLSy1 sources to host `young' jets.
\end{enumerate}

Bringing above mentioned points together, a promising scenario emerges where the \gm-NLSy1 galaxies can be understood as nascent blazars in which the jet activity was recently (in the cosmological context) triggered possibly due to galaxy mergers. Such interactions not only drive the gas inflows all the way to central engine but also can act as catalysts to spin-up the black hole. The 3D relativistic magnetohydrodynamic simulations indeed suggest that rapidly spinning black holes most efficiently produce jets (see, e.g., Tchekhovskoy {\em et al.} 2011, Tchekhovskoy \& McKinney 2012). Moreover, elliptical galaxies, which are likely to be the host of blazars, are thought to form via galaxy mergers (Toomre \& Toomre 1972; Di Matteo {\em et al.} 2005). Connecting this with mergers observed in \gm-NLSy1s, it can be understood that as the time progresses, late-type galaxies hosting \gm-NLSy1 may evolve to early-type ellipticals and a \gm-NLSy1 grows to become a powerful blazar. The following observed properties of \gm-NLSy1 objects provide further support to this scenario:

\begin{itemize}
\item The similarity of the flux variability behavior of the \gm-NLSy1s with blazars. In particular, rapid $\sim$hr scale flux variability is noted in the optical and \gm-ray bands similar to that observed from FSRQs.
\item The similarity of the broadband SED of the \gm-NLSy1 galaxies with FSRQs, e.g., a Compton dominated, LSP type SED. This indicates a similar environment surrounding the jet with dense photon fields originated from the luminous accretion disk, BLR, and torus contributing to the external Compton process (Figure~\ref{fig:SED}). The presence of a luminous BLR also explains the observation of strong (EW$>$5\AA) emission lines which, in turn, indicates an efficient and rapid accretion process similar to FSRQs.
\end{itemize}

To summarize, though the sample size of \gm-NLSy1s is currently too small to make any strong conclusions, their observed properties are intriguing and demands further exploration. The \gm-NLSy1 galaxies are probably the best systems to unravel the origin of the relativistic jet and to probe what is likely to be the most energetic and revolutionary phase in the galaxy evolution scheme.

\section*{Acknowledgements}

The author is grateful to referees for constructive suggestions and L. Foschini and J. Kotilainen for granting permission to use the host galaxy images of 1H 0323+342 and PKS 2004$-$447, respectively. Thanks are due to C. S. Stalin for a useful discussion. 

The \textit{Fermi} LAT Collaboration acknowledges generous ongoing support from a number of agencies and institutes that have supported both the development and the operation of the LAT as well as scientific data analysis. These include the National Aeronautics and Space Administration and the Department of Energy in the United States, the Commissariat \`a l'Energie Atomique and the Centre National de la Recherche Scientifique / Institut National de Physique Nucl\'eaire et de Physique des Particules in France, the Agenzia Spaziale Italiana and the Istituto Nazionale di Fisica Nucleare in Italy, the Ministry of Education, Culture, Sports, Science and Technology (MEXT), High Energy Accelerator Research Organization (KEK) and Japan Aerospace Exploration Agency (JAXA) in Japan, and the K.~A.~Wallenberg Foundation, the Swedish Research Council and the Swedish National Space Board in Sweden. Additional support for science analysis during the operations phase is gratefully acknowledged from the Istituto Nazionale di Astrofisica in Italy and the Centre National d'\'Etudes Spatiales in France. This work performed in part under DOE Contract DE- AC02-76SF00515.

The CSS survey is funded by the National Aeronautics and Space Administration under Grant No. NNG05GF22G issued through the Science Mission Directorate Near-Earth Objects Observations Program.  The CRTS survey is supported by the U.S.~National Science Foundation under grants AST-0909182 and AST-1313422.

This research has made use of data from the OVRO 40-m monitoring program (Richards {\em et al.} 2011) which is supported in part by NASA grants NNX08AW31G, NNX11A043G, and NNX14AQ89G and NSF grants AST-0808050 and AST-1109911.

Funding for the Sloan Digital Sky Survey IV has been provided by the Alfred P. Sloan Foundation, the U.S. Department of Energy Office of Science, and the Participating Institutions. SDSS-IV acknowledges support and resources from the Center for High-Performance Computing at the University of Utah. The SDSS web site is www.sdss.org. SDSS-IV is managed by the Astrophysical Research Consortium for the  Participating Institutions of the SDSS Collaboration including the Brazilian Participation Group, the Carnegie Institution for Science, Carnegie Mellon University, the Chilean Participation Group, the French Participation Group, Harvard-Smithsonian Center for Astrophysics, Instituto de Astrof\'isica de Canarias, The Johns Hopkins University, Kavli Institute for the Physics and Mathematics of the Universe (IPMU) / University of Tokyo, the Korean Participation Group, Lawrence Berkeley National Laboratory, Leibniz Institut f\"ur Astrophysik Potsdam (AIP), Max-Planck-Institut f\"ur Astronomie (MPIA Heidelberg), Max-Planck-Institut f\"ur Astrophysik (MPA Garching), Max-Planck-Institut f\"ur Extraterrestrische Physik (MPE), National Astronomical Observatories of China, New Mexico State University, New York University, University of Notre Dame, Observat\'ario Nacional / MCTI, The Ohio State University, Pennsylvania State University, Shanghai Astronomical Observatory, United Kingdom Participation Group, Universidad Nacional Aut\'onoma de M\'exico, University of Arizona, University of Colorado Boulder, University of Oxford, University of Portsmouth, University of Utah, University of Virginia, University of Washington, University of Wisconsin, Vanderbilt University, and Yale University.\vspace{-1em}

%%use \balance somewhere in the left column of the last page to balance the two columns in the end page

%%References section

\end{document}